\documentclass[prc,nofootinbib,showpacs]{revtex4}

\usepackage{graphicx}
\usepackage[usenames,dvipsnames]{color}
\usepackage{amsmath,amssymb,bbold}
\usepackage{hyperref}
\usepackage{comment}

\begin{document}

\title{Magneto-Seebeck coefficient and Nernst coefficient of hot and dense hadron gas}

\author{Arpan Das$^{1}$}
\email{arpan.das@ifj.edu.pl}
\author{Hiranmaya Mishra$^{2}$}
\email{hm@prl.res.in}
\author{Ranjita K. Mohapatra${^3}$}
\email{ranjita.iop@gmail.com}

\affiliation{$^{1}$Institute  of  Nuclear  Physics  Polish  Academy  of  Sciences,  PL-31-342  Krakow,  Poland}
\affiliation{$^{2}$Theory Division, Physical Research Laboratory, 
Navrangpura, Ahmedabad 380 009, India}
\affiliation{$^{3}$ Department of Physics, Banki Autonomous College, Cuttack 754008, India.}

\begin{abstract}
We discuss the thermoelectric effect of hot and dense hadron gas within the framework of the hadron resonance gas model. Using the relativistic Boltzmann equation within the relaxation time approximation we estimate the Seebeck coefficient of the hot and dense hadronic medium with a gradient in temperature and baryon chemical potential. The hadronic medium in this calculation is modeled by the hadron resonance gas (HRG) model with hadrons and their resonances up to a mass cutoff $\Lambda\sim 2.6$ GeV. We also extend the formalism of the thermoelectric effect for a nonvanishing magnetic field. The presence of magnetic field also leads to a Hall type thermoelectric coefficient (Nernst coefficient) for the hot and dense hadronic matter apart from a magneto-Seebeck coefficient. We find that generically in the presence of a magnetic field Seebeck coefficient decreases while the Nernst coefficient increases with the magnetic field. At higher temperature and/or baryon chemical potential these coefficients approach to their values at vanishing magnetic field.
\end{abstract}

\pacs{25.75.-q, 12.38.Mh}
\maketitle

\section{INTRODUCTION}

\label{intro}

Transport coefficients are important characteristics of thermodynamic systems that determine the evolution of the system towards equilibrium starting from an initial out of equilibrium state. A new state of strongly interacting matter has been reported in the relativistic heavy-ion collision experiments, e.g. at the Relativistic Heavy Ion Collider (RHIC) and the Large Hadron Collider (LHC). This strongly interacting matter so produced is also expected to achieve local thermal equilibrium within a few Fermi timescale after the collision, which is supported by the fact that relativistic hydrodynamical modeling of the strongly interacting medium successfully explains the particle spectra coming out of the medium after hadronization. Similar to other thermodynamic systems transport coefficients of the strongly interacting plasma are of utmost importance for a comprehensive understanding of Quantum Chromodynamics (QCD) in the nonperturbative regime. 
In the dissipative relativistic hydrodynamical model as well as in the transport simulations of the hot and dense medium these transport coefficients, e.g. shear and bulk viscosity, etc play important roles. It has been argued that the strongly interacting medium produced at RHIC is the most perfect fluid in nature having the smallest value of shear viscosity to entropy ratio ($\eta/s$) known to us.\cite{HeinzSnellings2013,RomatschkeRomatschke,KSS}. 
Apart from the shear viscosity another important transport coefficient which play an important role in the hydrodynamical evolution of non conformal field theories is the 
bulk viscosity $\zeta$ \cite{gavin1985,kajantie1985, DobadoTorres2012,sasakiRedlich2009,sasakiRedlich2010,KarschKharzeevTuchin2008,
FinazzoRougemont2015,WiranataPrakash2009,JeonYaffe1996}.
Although at asymptotically high temperature relative to $\Lambda_{\text{QCD}}$ strongly interacting plasma can be approximated with the conformal equation of state $(\mathcal{E}=3P)$,
lattice QCD simulations shows a nonmonotonic
behaviour of both $\eta/s$ and $\zeta/s$  near the  critical temperature $T_c$.  \cite{DobadoTorres2012,sasakiRedlich2009,sasakiRedlich2010,KarschKharzeevTuchin2008,
FinazzoRougemont2015,WiranataPrakash2009,JeonYaffe1996}.
It is important to note that the bulk viscosity encodes the conformal measure, $(\mathcal{E}-3P)/T^4$ of the system and nonvanishing bulk viscous coefficient indicates deviation from conformality of the strongly interacting matter. A new dimension has been added to the study of the strongly interacting plasma is the prediction of the generation of a strong magnetic field in the noncentral heavy-ion collisions. Conservative estimates indicate that the magnitude of the magnetic field can be as large as of several $m_{\pi}^2$, at least in the initial stage at RHIC energies \cite{mclerran2008,skokov}. The strength of the magnetic field in the initial stage can be large with respect to $\Lambda_{\text{QCD}}$, but in the absence of any thermalized conducting medium, this large magnetic field will quickly vanish. If the thermalization time of the strongly interacting medium is small then it can be argued that due to finite electrical conductivity some fraction of the initial large magnetic field may sustain in the conducting plasma \cite{MHD1,MHDajit,TuchinMHD,MoritzGreif,electricalcond1,electricalcond2,electricalcond3,electricalcond4,
electricalcond5,electricalcond6,electricalcond7,electricalcond8,electricalcond9,electricalcond10,electricalcond11,electricalcond12,
electricalcond13,electricalcond14,electricalcond15}.   Certainly, it is not known what is the exact amount of magnetic field that can be sustained in the medium, but taking an optimistic approach many investigations are going on to understand novel effects of magnetic field on QCD plasma. One of such effect is the CP-violating effects, e.g. chiral magnetic effect and chiral vortical effect  \cite{kharzeevbook}. The success of hydrodynamical modeling of the QCD plasma inspires a deeper understanding of hydrodynamical behavior in the presence of a magnetic field. In this context, magnetohydrodynamic simulations have been used to study the flow coefficient of the strongly interacting matter produced in heavy-ion collisions \cite{MHD1,MHDajit}. This apart various approaches like perturbative QCD (pQCD), and different QCD inspired effective models, etc have been used to estimate various transport coefficients for the QCD matter both in the absence and in the presence of magnetic field \cite{danicol2018,PrakashVenu,WiranataPrakash2012,
KapustaChakraborty2011,Toneev2010,Plumari2012,Gorenstein2008,Greiner2012,TiwariSrivastava2012,GhoshMajumder2013,Weise2015,GhoshSarkar2014,
WiranataKoch,WiranataPrakashChakrabarty2012,Wahba2010,Greiner2009,KadamHM2015,Kadam2015,Ghoshijmp2014,Demir2014,Ghosh2014,smash,bamps,bamps2,
urqmd1,GURUHM2015, ranjitahm,amanhm1,amanhm2,arpanhm,ranjitaarpanhm1,ranjitaarpanhm2,ranjitaarpanhm3,jayantadey1,feng2017}.
Another transport coefficient which becomes important in the presence of temperature gradient at finite baryon density is the thermal conductivity \cite{danicol2014,Kapusta2012}. The effect of the magnetic field on thermal conductivity of hot and dense hadron gas has been discussed in the literature \cite{ranjitaarpanhm2}.

In the present article, we investigate thermoelectric behavior in the presence of the magnetic field, specifically the magneto-Seebeck coefficient and Nernst coefficient of the hadronic medium produced in the subsequent evolution of the partonic medium formed in heavy-ion collisions. Due to nonvanishing thermoelectric coefficients, a temperature gradient in a conducting medium gets converted into an electrical current and vice versa. The  Seebeck coefficient is a measure of the electric field produced in a conducting medium due to a temperature gradient and it is formally defined by the ``open circuit'' condition i.e. when the electric current is set to zero \cite{callen,sofo}. In condensed matter systems Seebeck effect  has been investigated extensively e.g. in superconductors \cite{conseeb1,conseeb2}, graphene-superconductor junction \cite{conseeb3}, Majorana bound state coupled to a quantum dots \cite{conseeb4}, high temperature cuprates\cite{conseeb5}, superconductor-ferromagnetic tunnel junctions \cite{conseeb6}, low dimensional organic metals \cite{conseeb7}etc. Following the formalism developed in condensed matter systems as given in Refs.\cite{linearizedBoltzmann,thermoelectrics}, earlier some of us had studied the Seebeck coefficient in the absence of a magnetic field for the hot and dense hadronic matter within the framework of the hadron resonance gas model \cite{arpanhm}. The present investigation differs from our earlier work in the following ways: in Ref.\cite{arpanhm} we had investigated thermoelectric effect for a system that has a temperature gradient but we had considered spatially uniform chemical potential. Here we consider a temperature gradient as well as a gradient in the baryon chemical potential to estimate the thermoelectric coefficients. Due to Gibbs-Duhem relation, one can relate the temperature gradient with the gradient in the baryon chemical potential in the QCD medium \cite{gavin1985}. It is important to note that, contrary to a nonrelativistic system where heat current is defined with respect to Fermi surface, in a relativistic system thermal current can only be defined in the presence of conserved number current, e.g. for strongly interacting matter one generally defines heat current with respect to the baryon current due to the fact that baryon number is conserved in QCD. This is an important difference with respect to the definition of heat current in the nonrelativistic systems. These differences between the two formalisms are very important and due to these differences, the nature of estimated thermoelectric coefficients will be very different which we will show later in the formalism as well as in the result section. Further, we have extended the formalism of the thermoelectric effect for the nonvanishing magnetic field and estimate the magneto-Seebeck coefficient and the Nernst coefficient. Unlike the Seebeck coefficient, the Nernst coefficient is a Hall-type thermoelectric coefficient having nonvanishing value only for a finite magnetic field. If a conducting medium is subjected to a magnetic field in the presence of temperature gradient perpendicular to the magnetic field then the Nernst coefficient is the manifestation of electric current normal to both magnetic field and temperature gradient.

  In the condensed matter systems in the absence of a magnetic field nonvanishing Seebeck coefficient is ensured by the presence of temperature gradient only as because in these systems only one type of charge carriers are available e.g. either electrons or holes. Also for the electron-ion plasma where the different mobility of electrons and ions can give rise to a nonvanishing value of the Seebeck coefficient. On the other hand for an electron-positron plasma just having a temperature gradient does not guaranty any thermoelectric current in the absence of a magnetic field due to the cancellation of electric current due to particles and antiparticles. For strongly interacting plasma since heat current can only be defined with reference to the net baryon current, the Seebeck coefficient is only meaningful at finite baryon chemical potential.
 It is important to note that although the Seebeck coefficient is associated with electric current due to temperature gradient, due to Onsager relations thermoelectric coefficients associated with electrical current and the thermoelectric coefficients associated with the heat current are related, e.g. Peltier coefficient and Seebeck coefficient are not independent \cite{onsager1}. At finite baryon chemical potential, the number of baryons and anti-baryons are different and a net thermoelectric current driven by the temperature gradient can be produced. For heavy-ion collision experiments at RHIC and LHC the medium produced is expected to have very small net baryon number density, however, the heavy-ion collisions at Facility for Antiproton and Ion Research (FAIR) at Darmstadt \cite{fairref} and in Nuclotron-based Ion Collider fAcility (NICA) at Dubna \cite{nica} one expects a baryon-rich medium. In these low energy experiments, the thermalization of the strongly interacting medium is expected with finite baryon density. All the conditions for a nonvanishing thermoelectric effect, e.g. temperature gradient, nonvanishing baryon number, etc., allows us to study thermoelectric effects in these low energy heavy-ion collision experiments. Description of thermoelectric coefficients becomes more complicated in the presence of a magnetic field and in this investigation we have discussed in detail the effect of the magnetic field on the Seebeck coefficient and the Nernst coefficient. We have found that generically in the presence of magnetic field Seebeck coefficient of hot and dense hadron gas decreases. On the other hand, the Nernst coefficient increases with the magnetic field. Here we have assumed that in the hadronic system the strength of the magnetic field is small and it is not a dominant scale. 
 In the present approach, therefore, we attempted to estimate the transport coefficients where the phase space and the single-particle energies are not affected by the magnetic field through Landau quantization as in Refs.\cite{feng2017,ranjitaarpanhm1,ranjitaarpanhm2,ranjitaarpanhm3}. On the other hand, the effect of the magnetic field enters through the cyclotron frequency of the individual hadrons.

 In the context of heavy ion collision, the hadronic phase of the QCD plasma can be successfully described by the hadron resonance gas (HRG) model, at chemical freeze-out \cite{HRG1,HRG2,HRG3}. In the simplest approximation the strange and nonstrange particles freeze out in the same manner and in this simple case HRG model can be described by only two parameters i.e. temperature ($T$) and baryon chemical potential ($\mu_B$). It has been argued that thermodynamics of strongly interacting hadronic systems can be approximated by a system of noninteracting hadrons and its resonances, where the interactions between different hadrons are encoded as the resonances \cite{HRG4,HRG5}. HRG model has been explored regarding thermodynamics of hadronic medium \cite{thermodynamicsHRG1,thermodynamicsHRG2}, conserved charge fluctuations\cite{hrgfluc3,hrgfluc4,hrgfluc5,hrgfluc6,
hrgfluc7,ranjita2019} as well as transport coefficients for
hadronic matter\cite{MoritzGreif,PrakashVenu,WiranataPrakash2012,
KapustaChakraborty2011,Toneev2010,Gorenstein2008,Greiner2012,TiwariSrivastava2012,GhoshMajumder2013,Weise2015,GhoshSarkar2014,
WiranataKoch,Wahba2010,Greiner2009,KadamHM2015,Kadam2015,Ghoshijmp2014,Demir2014,Ghosh2014,smash,urqmd1,GURUHM2015}. Subsequently, the ideal HRG model has been improved upon, e.g. considering excluded volume due to the finite size of the hadrons   \cite{stockerRischke,GURUHM2015}. 
We would like to mention here that, although some of us studied the Seebeck coefficient for the hadronic medium using the formalism originally developed for condensed matter system, here we study the magneto-Seebeck coefficient and Nernst coefficient for the hadronic system using the formalism of thermoelectric effect compatible with relativistic systems. 

This paper is organized as follows, in Sec. \ref{formalism1} we introduce the formalism to estimate 
Seebeck coefficient in the absence of a magnetic field for a relativistic system which will be followed by the formalism of magneto-Seebeck coefficient and Nernst effect in Sec.\ref{formalism2}. In Sec. \ref{HRGmodel} we briefly discuss the HRG model and estimation of the relaxation time for hadrons and resonances.
In Sec. \ref{results} we present and discuss the results for the magneto-Seebeck coefficient and the Nernst coefficient. Finally, we summarize our work with an outlook in the
conclusion section.

\section{Boltzmann equation in relaxation time approximation and thermoelectric effect}
\label{formalism1}

 We first discuss here the thermoelectric effect in the absence of a magnetic field. We consider here the linearized Boltzmann equation in relaxation time approximation. The thermal equilibrium is achieved locally due to strong interaction and the external electromagnetic field acts as perturbation which takes the system slightly away from equilibrium.
 Under this approximation, the Boltzmann equation can be interpreted as a linear expansion of the distribution 
function around the equilibrium distribution function. $f(\vec{r},\vec{k})$ denotes out of equilibrium distribution function and the equilibrium distribution function is denoted as $f^{(0)}(\vec{r},\vec{k})$. Spatial dependence of the distribution functions appears due to spatial dependence of temperature $(T)$ and baryon chemical potential ($\mu_B$), in local thermodynamic equilibrium. For a global thermodynamic equilibrium temperature and chemical potential are uniform throughout the system hence for global thermodynamic equilibrium, distribution function does not depend on the coordinates rather it is the only function of momentum. For the linear response, we can write the linearized Boltzmann equation in the local rest frame, for particle species ``a'' in the following manner \cite{linearizedBoltzmann,thermoelectrics},

\begin{equation}
 \vec{v}_a.\vec{\nabla}f_a^{(0)}+q_a\vec{E}.\frac{\partial f_a^{(0)}}{\partial\vec{p}_a} = -\frac{f_a(\vec{r},\vec{p})-f_a^{(0)}(\vec{r},\vec{p})}{\tau_a(\vec{p})}
 \equiv -\frac{\delta f_a(\vec{r},\vec{p})}{\tau_a(\vec{p})} ,
 \label{equnew1}
\end{equation}
where $\delta f_a$ denotes the deviation from equilibrium and $\tau_a$ denotes the relaxation time of the particle species ``a''. The equilibrium distribution function satisfies, 
\begin{align}
 \frac{\partial f^{(0)}_{a}}{\partial \vec{p}_a}=\vec{v}_a\frac{\partial f^{(0)}_{a}}{\partial \epsilon_a},
 ~~ \frac{\partial f^{(0)}_{a}}{\partial \epsilon_a}=-\beta f^{(0)}_{a}(1\mp f^{(0)}_{a}), 
 ~~f^{(0)}_{a}=\frac{1}{e^{(\epsilon_a-b_a\mu_B)/T}\pm1},
\label{equnew2}
 \end{align}
 here the single particle energy is denoted as $\epsilon_a(p_a)=\sqrt{\vec{p}_a^2+m_a^2}$,  $\mu_B$ is the baryon chemical potential, $b_a$ denotes the baryon number of particle ``a'', e.g. for baryons $b_{\text{baryon}}=1$, for antibaryons $b_{\text{antibaryon}}=-1$ and for mesons $b_{\text{meson}}=0$. $\beta\equiv1/T$, is the 
 inverse of temperature. $\vec{v}_a=\vec{p}_a/\epsilon_a$ is the velocity of the particle. $\mp$
 is for fermion and boson respectively. The gradient of the equilibrium distribution function $\vec{\nabla}f^{(0)}_{a}$ can be expressed as,
\begin{align}
 \vec{\nabla}f^{(0)}_{a} = T \frac{\partial f^{(0)}_{a}}{\partial\epsilon_a}\bigg[\epsilon_a\vec{\nabla}\left(\frac{1}{T}\right)-b_a\vec{\nabla}\left(\frac{\mu_B}{T}\right)\bigg].
 \label{equnew3}
\end{align}

The spatial gradients of temperature and chemical potential can be related using
momentum conservation in the system and Gibbs Duhem relation. Momentum conservation leads to  $\partial_i P=0$. Using Gibbs Duhem relation we then have,
\begin{equation}
\partial _i P= \frac{\omega}{T}\partial_i T+T n_B\partial_i(\mu_B/T)=0.
\end{equation}
This relates the spatial gradient of temperature to the spatial gradients in chemical potential as,
\begin{equation}
\partial_i\mu_B=\left(\mu_B-\frac{\omega}{n_B}\right)\frac{\partial_i T}{T}.
\label{tmuder}
\end{equation}
Using Eq.(\ref{tmuder}) and Eq.\ref{equnew3}, we have
\begin{align}
 \vec{\nabla}f^{(0)}_{a} = - \frac{\partial f^{(0)}_{a}}{\partial\epsilon_a}\bigg(\epsilon_a-b_a\frac{\omega}{n_B}\bigg)\frac{\vec{\nabla}T}{T}.
\label{equnew4}
 \end{align}
Using Eq.\eqref{equnew4} in Eq.\eqref{equnew1} we can write the deviation of the equilibrium distribution function as,
\begin{align}
 \delta f_a = -\tau_a\frac{\partial f_a^{(0)}}{\partial \epsilon_a}\bigg[q_a(\vec{E}.\vec{v}_a)-\bigg(\epsilon_a-b_a\frac{\omega}{n_B}\bigg)\vec{v}_a.\vec{\nabla}\ln T\bigg].
 \label{equnew5}
\end{align}
Here we have written, $\vec{\nabla} \ln T = \frac{\vec{\nabla} T}{T}$.
Using the deviation from the equilibrium distribution function as given in Eq.\eqref{equnew5}, the electric current $(\vec{j})$ of the system can be defined as,
\begin{align}
 \vec{j}& = \sum_a g_a\int \frac{d^3p_a}{(2\pi)^3}q_a\vec{v}_a\delta f_a\nonumber\\
 & = \sum_a \frac{g_a}{3} \int \frac{d^3p_a}{(2\pi)^3}\tau_a q_a^2 v^2_a\bigg(-\frac{\partial f_a^{(0)}}{\partial \epsilon_a}\bigg)\vec{E}\nonumber\\
 & ~~~~~-\sum_a \frac{g_a}{3} \int \frac{d^3p_a}{(2\pi)^3}\tau_aq_av^2_a
 \bigg(\epsilon_a-b_a\frac{\omega}{n_B}\bigg)\bigg(-\frac{\partial f_a^{(0)}}{\partial \epsilon_a}\bigg)\vec{\nabla}\ln T.
 \label{equnew6}
\end{align}
In Eq.\eqref{equnew6} we have used $\langle v^i_av^j_a\rangle=\frac{1}{3}v_a^2\delta^{ij}$. Further, the sum is over all the baryons and mesons including their antiparticles. 
For a relativistic system thermal current or the heat current arises corresponding to a conserved particle number. Thermal conduction is defined with reference to the conserved baryon current.  
Thermal conduction arises when energy flows relative to the baryonic enthalpy. 
Hence in the presence of conserved baryon current, the heat current of hadron resonance gas can be defined as \cite{gavin1985} ,

\begin{align}
 \mathcal{I}^i = \sum_{a} T_a^{0i}-\frac{\omega}{n_B}\sum_{a}b_a j_{B_a}^i.
 \label{equnew7}
\end{align}
In Eq.\eqref{equnew7} $\omega = \mathcal{E}+P$ is the enthalpy of the system, $\mathcal{E}$ is the energy density and $P$ is the pressure of the system. In Eq.\eqref{equnew7} $T^{0i}$ is the ``0i-th'' component of the energy momentum tensor $T^{\mu\nu}$. Conserved baryon current is denoted as $j_{B}^{\mu}$ and $n_B$ represents the net baryon number density. Using the definition of energy momentum tensor ($T^{\mu\nu}$) and baryon current ($j_B^{\mu}$), heat current $\mathcal{I}^i$ as given in Eq.\eqref{equnew7} can be recasted as \cite{gavin1985}, 
\begin{align}
 \mathcal{I}^i & = \sum_a g_a\int \frac{d^3p_a}{(2\pi)^3}p_a^if_a -\frac{\omega}{n_B} \sum_a b_a g_a\int\frac{d^3p_a}{(2\pi)^3} v_a^i f_a\nonumber\\
 & = \sum_a g_a\int \frac{d^3p_a}{(2\pi)^3}\frac{p_a^i}{\epsilon_a}\left(\epsilon_a-b_a\frac{\omega}{n_B}\right)\delta f_a.
 \label{equnew8}
\end{align}
In Eq.\eqref{equnew8} equilibrium distribution function does not contributes because the equilibrium distribution function is a even function of momentum.  Using the expression of $\delta f_a$ as given in Eq.\eqref{equnew5} in Eq.\eqref{equnew8} we get for the heat current,
\begin{align}
 \mathcal{I}^i
 & = \sum_a g_a\int \frac{d^3p_a}{(2\pi)^3}\frac{p_a^i}{\epsilon_a}\left(\epsilon_a-b_a\frac{\omega}{n_B}\right)\delta f_a\nonumber\\
 & = \sum_a \frac{g_a}{3} \int \frac{d^3p_a}{(2\pi)^3} \tau_a q_a v_a^2\left(\epsilon_a-b_a\frac{\omega}{n_B}\right)\left(-\frac{\partial f_a^{(0)}}{\partial\epsilon_a}\right)\vec{E}\nonumber\\
 & ~~~~~-\sum_a \frac{g_a}{3} \int \frac{d^3p_a}{(2\pi)^3} \tau_a v_a^2\left(\epsilon_a-b_a\frac{\omega}{n_B}\right)^2\left(-\frac{\partial f_a^{(0)}}{\partial\epsilon_a}\right)\vec{\nabla}\ln T.
 \label{equnew9}
\end{align}
The Seebeck coefficient $S$ can be  determined using Eq.\eqref{equnew6} by setting $\vec{j}=0$, so that the electric field becomes proportional to the temperature gradient and the proportionality factor is nothing but the Seebeck coefficient \cite{linearizedBoltzmann,thermoelectrics}. Hence from  Eq.\eqref{equnew6} we get, 
\begin{align}
 \vec{E}=S\vec{\nabla}T,
\end{align}
here the Seebeck coefficient in the Boltzmann approximation can be identified as, 
\begin{align}
 S &= \frac{\sum_a \frac{g_a}{3}\int \frac{d^3p_a}{(2\pi)^3}\tau_a q_a v_a^2\left(\epsilon_a-b_a\frac{\omega}{n_B}\right)\left(-\frac{\partial f_a^{(0)}}{\partial\epsilon_a}\right)}{T\sum_a \frac{g_a}{3}\int \frac{d^3p_a}{(2\pi)^3}\tau_a q^2_a v_a^2\left(-\frac{\partial f_a^{(0)}}{\partial\epsilon_a}\right)}\nonumber\\
 &=\frac{\sum_a \frac{g_a}{3T}\int \frac{d^3p_a}{(2\pi)^3}\tau_a q_a \left(\frac{\vec{p_a}}{\epsilon_a}\right)^2\left(\epsilon_a-b_a\frac{\omega}{n_B}\right)f_a^{(0)}}{T\sum_a \frac{g_a}{3T}\int \frac{d^3p_a}{(2\pi)^3}\tau_a q^2_a \left(\frac{\vec{p_a}}{\epsilon_a}\right)^2f_a^{(0)}}\equiv \frac{\mathcal{I}_{31}/T^2}{\sigma_{el}/T}.
 \label{equnew11}
\end{align}
Let us note that at small temperature and large enough baryon chemical potential for which $Ts<<\mu_B n_B$, $\left(\epsilon_a-b_a\frac{\omega}{n_B}\right) \simeq(\epsilon_a-b_a \mu_B)$ and in this limit the Seebeck coefficient reduces to the expression for the same given in Ref.\cite{arpanhm}. However it may be noted for the system of hadron resonance gas the contribution of pions to entropy can be large and it should not be ignored. We will discuss it more in Sec.\eqref{results}.
The electrical conductivity $\sigma_{el}$ can be identified from Eq.\eqref{equnew6} as,
\begin{align}
 \sigma_{el}  & = \sum_a \frac{g_a}{3T}\int \frac{d^3p_a}{(2\pi)^3}\tau_a q^2_a \left(\frac{\vec{p}_a}{\epsilon_a}\right)^2f_a^{(0)},
\end{align}
and the integral $\mathcal{I}_{31}$ in Eq.\eqref{equnew11} is, 
\begin{align}
 \mathcal{I}_{31} = \sum_a \frac{g_a}{3T}\int \frac{d^3p_a}{(2\pi)^3}\tau_a q_a \left(\frac{\vec{p_a}}{\epsilon_a}\right)^2\left(\epsilon_a-b_a\frac{\omega}{n_B}\right)f_a^{(0)}. 
 \label{I31equ}
\end{align}
We would like to make two comments regarding the expression for the Seebeck coefficient as given in Eq.\eqref{equnew11}. Firstly, Seebeck coefficient can be positive and negative as the numerator depends linearly on electric charge while the integrand itself is not manifestly positive definite. Secondly in the numerator only the baryons will contribute as the mesonic contribution will cancel out. In terms  of $\sigma_{el}$ and $S$ the electric current can be expressed  as,
\begin{equation}
 \vec{j}=\sigma_{el}\vec{E}-\sigma_{el}S \vec{\nabla}T.
 \label{equnew13}
\end{equation}
In a similar way, the heat current as given in Eq.\eqref{equnew9} can be expressed as,
\begin{equation}
 \vec{\mathcal{I}}=T\sigma_{el}S\vec{E}-k_0\vec{\nabla}T,
 \label{equnew14}
\end{equation}
where $k_0$ is the thermal conductivity and is expressed as \cite{gavin1985},
\begin{align}
 k_0=\sum_a\frac{g_a}{3T^2}\int\frac{d^3p_a}{(2\pi)^3}\tau_a\left(\frac{\vec{p}_a}{\epsilon_a}\right)^2\left(\epsilon_a-b_a\frac{\omega}{n_B}\right)^2f_a^{(0)}.
 \label{equnew15}
\end{align}
Using Eq.\eqref{equnew13} and Eq.\eqref{equnew14}, we can express the heat current $\vec{\mathcal{I}}$ in terms  of electric current $\vec{j}$ in the following way,
\begin{equation}
 \vec{\mathcal{I}}=TS\vec{j}-\left(k_0-T\sigma_{el}S^2\right)\vec{\nabla}T.
 \label{equnew16}
\end{equation}
From Eq.\eqref{equnew16} we can identify the Peltier coefficient $(\Pi)$ and thermal conductivity $(k)$ in the presence of non vanishing Seebeck coefficient as,
\begin{align}
 & \Pi=TS,\label{equnew17}\\
 & k = k_0-T\sigma_{el}S^2\label{equnew18}.
\end{align}

Note that the Peltier coefficient ($\Pi$) in terms of the Seebeck coefficient ($S$) as given in Eq.\eqref{equnew17} can be considered as the consistency relation which can also be obtained using Onsager relation \cite{onsager1}. Also, note that from Eq.\eqref{equnew18} the thermal conductivity in the absence of any thermoelectric effect matches the expression of the thermal conductivity as given in Ref.\cite{gavin1985}.
Here some comments on the formalism as given earlier in Ref.\cite{arpanhm} is in order. The basic difference between the expression of the total Seebeck coefficient as given in Eq.\eqref{equnew11} for a multi-component system and the same as given in Ref.\cite{arpanhm}, is the presence of the factor $\omega/n_B$. In Ref.\cite{arpanhm} instead of the factor $\omega/n_B$, baryon chemical potential $\mu_B$ enters into the expression of the Seebeck coefficient. This is because in Ref.\cite{arpanhm} the spatial dependence of $\mu_B$ was ignored and the heat current is defined with respect to the baryon chemical potential rather that the conserved baryon current. However similar to Ref.\cite{arpanhm}, here also the mesonic contributions to the total Seebeck coefficient cancel out in the numerator of the Eq.\eqref{equnew11} due to the equal and opposite contribution of the particles and antiparticles. Only baryonic contribution becomes relevant in the numerator at finite $\mu_B$. Mesons only contribute to the total enthalpy of the system and also in the total electrical conductivity of the system which enters the denominator of the Eq.\eqref{equnew11}. This simple picture becomes more complicated in the presence of a magnetic field which we discuss in the next section.

\section{Magneto-Seebeck coefficient and Nernst coefficient}
\label{formalism2}
To investigate the effect of the magnetic field on the thermoelectric effect we start with the relativistic Boltzmann transport equation (RBTE) to estimate the deviation from equilibrium $\delta f^a$ in the presence of a magnetic field. The RBTE in presence of an electric field ($\vec{E}$) and a magnetic field ($\vec{B}$) for a single hadron species can be expressed as, 
\begin{align}
 \vec{v}_a.\frac{\partial f_a}{\partial \vec{r}}+q_a\left(\vec{E}+\vec{v}_a\times\vec{B}\right).\frac{\partial f_a}{\partial\vec{p}_a} = \mathcal{C}[f_a],
 \label{equnew19}
\end{align}
here in Eq.\eqref{equnew19} $q_a$ is the electric charge of the particle ``a'' and the collision integral is denoted as $\mathcal{C}[f_a]$. In global thermal equilibrium Eq.\eqref{equnew19} is trivially satisfied as both L.H.S and R.H.S in Eq.\eqref{equnew19} vanishes. Hence, we can express the Boltzmann kinetic equation as given in Eq.\eqref{equnew19} as an equation for the deviation from equilibrium distribution function, $\delta f_a =f_a-f^{(0)}_{a}$,
\begin{align}
 \vec{v}_a.\frac{\partial f^{(0)}_{a}}{\partial \vec{r}}+q_a\vec{E}.\frac{\partial f^{(0)}_{a}}{\partial\vec{p}}+q_a\left(\vec{v_a}\times\vec{B}\right).\frac{\partial (\delta f_a)}{\partial\vec{p_a}} = \mathcal{C}[\delta f_a].
\end{align}
Note that equilibrium distribution function $f^{(0)}_{a}$ does not contribute to the Lorentz force as because $\frac{\partial f^{(0)}_{a}}{\partial\vec{p}}\propto \vec{v}_a$, hence $\left(\vec{v_a}\times\vec{B}\right).\frac{\partial f^{(0)}_a}{\partial\vec{p_a}}=0$. Otherwise complicated collision integral $\mathcal{C}[f_a]$ takes a simple form in the relaxation time ($\tau_a$) approximation (RTA). In relaxation time approximation (RTA) $\mathcal{C}[\delta f_a]\equiv \delta f_a/\tau_a$. Therefore the RBTE in the relaxation time approximation get reduced to, 
\begin{align}
 \vec{v}_a.\frac{\partial f^{(0)}_{a}}{\partial \vec{r}}+q_a\vec{E}.\frac{\partial f^{(0)}_{a}}{\partial\vec{p}}+q_a\left(\vec{v_a}\times\vec{B}\right).\frac{\partial (\delta f_a)}{\partial\vec{p_a}} =  -\frac{\delta f_a}{\tau_a},
 \label{equnew21}
\end{align}
Some comment on the relaxation time approximation in the presence of a magnetic field is in order here. In the relaxation time approximation, the system is not far away from equilibrium due to external perturbation and then it relaxes back towards equilibrium with the characteristics time scale given as the relaxation time. Hence external perturbation is not the dominant scale. For the strongly interacting medium, the strong interaction is responsible for
equilibration of the system and the magnetic field is a perturbation with respect to the strong interaction.
In the heavy-ion collision, a large magnetic field can be produced at the initial stages. Conservative estimates predict the strength of the magnetic field is of the order of a few $m_{\pi}^2\sim0.02$ GeV$^2$ at RHIC energies and even less in the low energy collisions. However the magnetic field decreases rapidly in the vacuum, but in a conducting medium, the magnetic field can sustain for a larger time scale with respect to the vacuum. A small fraction of the initial magnetic field can survive in the hadronic phase. The relaxation time approximation can be justified in the presence of a weak magnetic field when the magnetic field is the subdominant scale. In the relaxation time approximation, the effect of the magnetic field has been ignored in the collision integral as well as in thermodynamics. But if the magnetic field is dominant then one has to carefully consider the effect of the magnetic field in the collision integral as well as in thermodynamics of the system.
To solve the RBTE as given in Eq.\eqref{equnew21} we take an ansatz to express the deviation of the distribution function from the equilibrium in the following way,
  \begin{align}
  \delta f_a = (\vec{p}_a.~\vec{\Xi})\frac{\partial f^{(0)}_{a}}{\partial\epsilon_a},
\label{equnew22}
  \end{align}
with $\vec{\Xi}$ being related to temperature gradient, electric field, the magnetic field and in general can be written as,
\begin{align}
 \vec{\Xi}= \alpha \vec{e}+\beta\vec{h}+\gamma(\vec{e}\times\vec{h})+\rho \vec{\nabla}T+ b(\vec{\nabla}T\times\vec{h}) +f (\vec{\nabla}T\times \vec{e}).
 \label{equnew23}
\end{align}
In Eq.\eqref{equnew23}, $\vec{h}=\frac{\vec{B}}{|B|}$ and $\vec{e}=\frac{\vec{E}}{|E|}$, is the direction of the magnetic field and electric field respectively. In the absence of a temperature gradient such an ansatz had been taken earlier in Ref.\cite{feng2017,ranjitaarpanhm1,ranjitaarpanhm2,ranjitaarpanhm3} which we have generalized to include a temperature gradient. Using Eq.\eqref{equnew22} and \eqref{equnew4},  RBTE as given in  Eq.\eqref{equnew21} can be expressed as,
\begin{align}
\vec{v}_a.\bigg[- \frac{\partial f^{(0)}_{a}}{\partial\epsilon_a}\bigg(\epsilon_a-b_a\frac{\omega}{n_B}\bigg)\frac{\vec{\nabla}T}{T}\bigg]+q_a(\vec{E}.\vec{v}_a)\frac{\partial f^{(0)}_{a}}{\partial \epsilon_a}-q_aB \vec{v}_a.(\vec{\Xi}\times\vec{h})\frac{\partial f^{(0)}_{a}}{\partial \epsilon_a} = -\frac{\epsilon_a}{\tau_a}(\vec{v}_a.\vec{\Xi})\frac{\partial f^{(0)}_{a}}{\partial \epsilon_a}.
\label{equnew24}
\end{align}
 One should note that in the presence of a magnetic field first law of thermodynamics as well as Gibbs-Duhem 
relation gets modified. However this modification involves non vanishing magnetization of the system. In this present investigation we are not considering spin-magnetic field interaction and magnetization of the system, hence we are considering Eq. \eqref{equnew4} even for a nonvanishing magnetic field.
Using the representation of $\vec{\Xi}$ as given in Eq.\eqref{equnew23}, RBTE as given in Eq.\eqref{equnew24}, can be expressed as,
\begin{align}
 &q_a(\vec{E}.\vec{v}_a)-\alpha q_aB\vec{v}_a.(\vec{e}\times\vec{h})-\gamma q_a B (\vec{e}.\vec{h})(\vec{v}_a.\vec{h})+\gamma q_a B(\vec{v}_a.\vec{e})-\rho q_a B \vec{v}_a.(\vec{\nabla}T\times \vec{h})-bq_aB(\vec{\nabla}T.\vec{h})(\vec{v}_a.\vec{h})\nonumber\\
 &~~~+bq_aB(\vec{v}_a.\vec{\nabla}T)-fq_aB(\vec{\nabla}T.\vec{h})(\vec{v}_a.\vec{e})+fq_aB(\vec{e}.\vec{h})(\vec{v}_a.\vec{\nabla}T)-\bigg(\epsilon_a-b_a\frac{\omega}{n_B}\bigg)(\vec{v}_a.\vec{\nabla}\ln T)\nonumber\\
 & = -\frac{\epsilon_a}{\tau_a}\bigg[\alpha(\vec{v}_a.\vec{e})+\beta(\vec{v}_a.\vec{h})+\gamma \vec{v}_a.(\vec{e}\times\vec{h})+\rho(\vec{v}_a.\vec{\nabla}T)+b\vec{v}_a.(\vec{\nabla}T\times \vec{h})+f\vec{v}_a.(\vec{\nabla}T\times\vec{e})\bigg].
 \label{equnew25}
\end{align}
Comparing the coefficients of different tensor structures on both sides of Eq.\eqref{equnew25} we get,
\begin{align}
 & f=0,\\
 & q_aE+\gamma q_aB =-\frac{\epsilon_a}{\tau_a}\alpha,\label{equnew27}\\
 & \gamma = \alpha \tau_a \omega_{c_a},\label{equnew28}\\
 & \beta = \gamma \tau_a \omega_{c_a}(\vec{e}.\vec{h})+b\tau_a\omega_{c_a}(\vec{\nabla}T.\vec{h}),\label{equnew29}\\
 & b= \omega_{c_a}\tau_a\rho,\label{equnew30}\\
 & \rho =\frac{\tau_a}{\epsilon_a}\frac{1}{T}\bigg(\epsilon_a-b_a\frac{\omega}{n_B}\bigg)-b \tau_a\omega_{c_a}\label{equnew31}.
\end{align}
Here $\omega_{c_a}=\frac{q_aB}{\epsilon_a}$ represents the cyclotron frequency of the particle with electric charge $q_a$. Using Eqs.\eqref{equnew27},\eqref{equnew28},\eqref{equnew29},\eqref{equnew30} and Eq.\eqref{equnew31}  it can be shown that,
\begin{align}
 & \alpha = -\frac{(q_aE)(\tau_a/\epsilon_a)}{1+(\omega_{c_a}\tau_a)^2},\label{equnew33}\\
 & \rho = \frac{\tau_a}{\epsilon_a} \frac{\bigg(\epsilon_a- b_a\frac{\omega}{n_B}\bigg)}{T}\frac{1}{1+(\omega_{c_a}\tau_a)^2}\label{equnew34}.
\end{align}
$\alpha$ and $\rho$ as given in Eq.\eqref{equnew33} and Eq.\eqref{equnew34} allows us to write the 
deviation from the equilibrium distribution function as,
\begin{align}
 \delta f_a = \frac{\tau_a}{1+(\omega_{c_a}\tau_a)^2}& \bigg[q_a\bigg\{(\vec{v}_a.\vec{E})+(\omega_{c_a}\tau_a)\vec{v}_a.(\vec{E}\times\vec{h})+(\omega_{c_a}\tau_a)^2(\vec{E}.\vec{h})(\vec{v}_a.\vec{h})\bigg\}\nonumber\\
 & -\frac{\bigg(\epsilon_a- b_a\frac{\omega}{n_B}\bigg)}{T}\bigg\{(\vec{v}_a.\vec{\nabla}T)+(\omega_{c_a}\tau_a)\vec{v}_a.(\vec{\nabla T\times h})+(\omega_{c_a}\tau_a)^2(\vec{\nabla}T.\vec{h})(\vec{v}_a.\vec{h})\bigg](-)\frac{\partial f_{0_a}}{\partial\epsilon_a}.
 \label{equnew35}
 \end{align}
 Using $\delta f^a$ as given in Eq.\eqref{equnew35} we can express the electrical current $(\vec{j})$ and the heat current $(\vec{\mathcal{I}})$ as,
 \begin{align}
  j^l & =\sum_a g_a \int \frac{d^3p_a}{(2\pi)^3}q_a v_a^{l}\delta f_a\nonumber\\
  & = \sum_a \frac{g_aq_a}{3}\int\frac{d^3p_a}{(2\pi)^3}\frac{v_a^2 \tau_a}{1+(\omega_{c_a}\tau_a)^2}\bigg[ q_a\delta^{lj}E^j+q_a (\omega_{c_a}\tau_a)\epsilon^{ljk}h^kE^j+q_a(\omega_{c_a}\tau_a)^2h^lh^jE^j\nonumber\\
  &~~~~~~~~~~~~~~~~~~~~~~~~~~~~~~ -\bigg(\epsilon_a- b_a\frac{\omega}{n_B}\bigg)\bigg\{\delta^{lj}\frac{\partial\ln T}{\partial x^j}+(\omega_{c_a}\tau_a)\epsilon^{ljk}h^k \frac{\partial\ln T}{\partial x^j}+(\omega_{c_a}\tau_a)^2h^lh^j\frac{\partial\ln T}{\partial x^j}\bigg\}\bigg](-)\frac{\partial f_a^{(0)}}{\partial \epsilon_a},
  \label{equnew36}
 \end{align}
and,
\begin{align}
 \mathcal{I}^l &  = \sum_a g_a \int \frac{d^3p_a}{(2\pi)^3}v_a^l\bigg(\epsilon_a-b_a\frac{\omega}{n_B}\bigg)\delta f_a\nonumber\\
 & = \sum_a \frac{g_a}{3}\int\frac{d^3p_a}{(2\pi)^3}\frac{v_a^2\tau_a}{1+(\omega_{c_a}\tau_a)^2}\bigg(\epsilon_a-b_a\frac{\omega}{n_B}\bigg)\bigg[ q_a\delta^{lj}E^j+q_a (\omega_{c_a}\tau_a)\epsilon^{ljk}h^kE^j+q_a(\omega_{c_a}\tau_a)^2h^lh^jE^j\nonumber\\
  &~~~~~~~~~~~~~~~~~~~~~~~~~~~~~~ -\bigg(\epsilon_a- b_a\frac{\omega}{n_B}\bigg)\bigg\{\delta^{lj}\frac{\partial\ln T}{\partial x^j}+(\omega_{c_a}\tau_a)\epsilon^{ljk}h^k \frac{\partial\ln T}{\partial x^j}+(\omega_{c_a}\tau_a)^2h^lh^j\frac{\partial\ln T}{\partial x^j}\bigg\}\bigg](-)\frac{\partial f_a^{(0)}}{\partial \epsilon_a}.
  \label{equnew37}
\end{align}
Electric current ($\vec{j}$) and the heat current ($\vec{\mathcal{I}}$) as given in Eq.\eqref{equnew36} and
Eq.\eqref{equnew37} respectively are quite general. However it is difficult to identify the thermoelectric coefficients from Eq.\eqref{equnew36} and Eq.\eqref{equnew37}. To simplify further calculation, without the loss of generality, we can choose the magnetic field along the $z$ direction and take the electric field ($\vec{E}$), and the temperature gradient ($\vec{\nabla}T$) perpendicular to it i.e.
in the $x-y$ plane. Under these conditions the components of the electric current in the $x-y$ plane are given as,
\begin{align}
 j_x = &  \sum_a\frac{g_aq_a}{3}\int\frac{d^3p_a}{(2\pi)^3}\frac{v_a^2q_a\tau_a}{1+(\omega_{c_a}\tau_a)^2}\bigg[E_x+(\omega_{c_a}\tau_a)E_y\bigg](-)\frac{\partial f_a^{(0)}}{\partial \epsilon_a}\nonumber\\
 & -\sum_a\frac{g_a q_a}{3T}\int\frac{d^3p_a}{(2\pi)^3}\frac{v_a^2\tau_a\left(\epsilon_a-b_a\frac{\omega}{n_B}\right)}{1+(\omega_{c_a}\tau)^2}\bigg[\frac{dT}{dx}+(\omega_{c_a}\tau_a)\frac{dT}{dy}\bigg](-)\frac{\partial f_a^{(0)}}{\partial \epsilon_a},
 \label{equnew38}
\end{align}
and,
\begin{align}
 j_y = &  \sum_a\frac{g_aq_a}{3}\int\frac{d^3p_a}{(2\pi)^3}\frac{v_a^2q_a\tau_a}{1+(\omega_{c_a}\tau_a)^2}\bigg[E_y-(\omega_{c_a}\tau_a)E_x\bigg](-)\frac{\partial f_a^{(0)}}{\partial \epsilon_a}\nonumber\\
 & -\sum_a\frac{g_a q_a}{3T}\int\frac{d^3p_a}{(2\pi)^3}\frac{v_a^2\tau_a\left(\epsilon_a-b_a\frac{\omega}{n_B}\right)}{1+(\omega_{c_a}\tau)^2}\bigg[\frac{dT}{dy}-(\omega_{c_a}\tau_a)\frac{dT}{dx}\bigg](-)\frac{\partial f_a^{(0)}}{\partial \epsilon_a}.
 \label{equnew39}
\end{align}
Eq.\eqref{equnew38} and \eqref{equnew39} can be written in a compact form by introducing the following integrals,
\begin{align}
 & L_{1_a}=\frac{g_a}{3}\int\frac{d^3p_a}{(2\pi)^3}\frac{\tau_a}{1+(\omega_{c_a}\tau_a)^2}\left(\frac{\vec{p}_a^2}{\epsilon_a^2}\right)(-)\frac{\partial f_a^{(0)}}{\partial \epsilon_a},\label{equnew40}\\
& L_{2_a}=\frac{g_a}{3}\int\frac{d^3p_a}{(2\pi)^3}\frac{\tau_a(\omega_{c_a}\tau_a)}{1+(\omega_{c_a}\tau_a)^2}\left(\frac{\vec{p}_a^2}{\epsilon_a^2}\right)(-)\frac{\partial f_a^{(0)}}{\partial \epsilon_a},\label{equnew41}\\
& L_{3_a}=\frac{g_a}{3}\int\frac{d^3p_a}{(2\pi)^3}\frac{\tau_a\epsilon_a}{1+(\omega_{c_a}\tau_a)^2}\left(\frac{\vec{p}_a^2}{\epsilon_a^2}\right)(-)\frac{\partial f_a^{(0)}}{\partial \epsilon_a},\label{equnew42}\\
& L_{4_a}=\frac{g_a}{3}\int\frac{d^3p_a}{(2\pi)^3}\frac{\tau_a\epsilon_a(\omega_{c_a}\tau_a)}{1+(\omega_{c_a}\tau_a)^2}\left(\frac{\vec{p}_a^2}{\epsilon_a^2}\right)(-)\frac{\partial f_a^{(0)}}{\partial \epsilon_a}.\label{equnew43}
 \end{align}
The integrals as given in Eq.\eqref{equnew40}-Eq.\eqref{equnew43} allows us to write  Eq.\eqref{equnew38} and  Eq.\eqref{equnew39}, respectively, as 
\begin{align}
 j_x = \sum_a q_a^2L_{1_a}E_x+\sum_a q_a^2L_{2_a}E_y -\sum_{a}q_a\bigg(L_{3_a}-b_a\frac{\omega}{n_B}L_{1_a}\bigg)\frac{d\ln T}{dx}-\sum_{a}q_a\bigg(L_{4_a}-b_a\frac{\omega}{n_B}L_{2_a}\bigg)\frac{d\ln T}{dy},
 \label{equnew44}
\end{align}
and,
\begin{align}
 j_y = \sum_a q_a^2L_{1_a}E_y-\sum_a q_a^2L_{2_a}E_x -\sum_{a}q_a\bigg(L_{3_a}-b_a\frac{\omega}{n_B}L_{1_a}\bigg)\frac{d\ln T}{dy}+\sum_{a}q_a\bigg(L_{4_a}-b_a\frac{\omega}{n_B}L_{2_a}\bigg)\frac{d\ln T}{dx}.
 \label{equnew45}
\end{align}
Seebeck coefficient in the presence of magnetic field or the magneto-Seebeck coefficient ($S_B$) in this case can be  determined by setting $j_x=0$ and $j_y=0$, so that the electric field becomes proportional to the temperature gradient. For $j_x=0$ and $j_y=0$ we can solve Eq.\eqref{equnew44} and \eqref{equnew45} to get $E_x$ and $E_y$ in terms of temperature gradients $\frac{dT}{dx}$ and  $\frac{dT}{dy}$ in the following way,
\begin{align}
 E_x &  = \frac{\sum_a q_a^2L_{1_a}\sum_aq_a(L_{3_a}-b_a\frac{\omega}{n_B}L_{1_a})+\sum_a q_a^2L_{2_a}\sum_aq_a(L_{4_a}-b_a\frac{\omega}{n_B}L_{2_a})}{T\bigg[\bigg(\sum_a q_a^2L_{1_a}\bigg)^2+\bigg(\sum_a q_a^2L_{2_a}\bigg)^2\bigg]}\frac{dT}{dx}\nonumber\\
 & +\frac{\sum_a q_a^2L_{1_a}\sum_aq_a(L_{4_a}-b_a\frac{\omega}{n_B}L_{2_a})-\sum_a q_a^2L_{2_a}\sum_aq_a(L_{3_a}-b_a\frac{\omega}{n_B}L_{1_a})}{T\bigg[\bigg(\sum_a q_a^2L_{1_a}\bigg)^2+\bigg(\sum_a q_a^2L_{2_a}\bigg)^2\bigg]}\frac{dT}{dy},
 \label{equnew46}
\end{align}
and,
\begin{align}
 E_y &  = \frac{\sum_a q_a^2L_{2_a}\sum_aq_a(L_{3_a}-b_a\frac{\omega}{n_B}L_{1_a})-\sum_a q_a^2L_{1_a}\sum_aq_a(L_{4_a}-b_a\frac{\omega}{n_B}L_{2_a})}{T\bigg[\bigg(\sum_a q_a^2L_{1_a}\bigg)^2+\bigg(\sum_a q_a^2L_{2_a}\bigg)^2\bigg]}\frac{dT}{dx}\nonumber\\
 & +\frac{\sum_a q_a^2L_{1_a}\sum_aq_a(L_{3_a}-b_a\frac{\omega}{n_B}L_{1_a})+\sum_a q_a^2L_{2_a}\sum_aq_a(L_{4_a}-b_a\frac{\omega}{n_B}L_{2_a})}{T\bigg[\bigg(\sum_a q_a^2L_{1_a}\bigg)^2+\bigg(\sum_a q_a^2L_{2_a}\bigg)^2\bigg]}\frac{dT}{dy}.
 \label{equnew47}
\end{align}
Eq.\eqref{equnew46} and \eqref{equnew47} can be written in a compact form in the following way,
\begin{align}
 \begin{pmatrix}
E_x \\
\\
E_y 
\end{pmatrix}= \begin{pmatrix}
S_B & NB \\
\\
-NB & S_B 
\end{pmatrix}\begin{pmatrix}
\frac{dT}{dx} \\
\\
\frac{dT}{dy} 
\end{pmatrix},
\end{align}
here one can identify the magneto-Seebeck coefficient ($S_B$) as, 
\begin{align}
 S_B & = \frac{\sum_a q_a^2L_{1_a}\sum_aq_a(L_{3_a}-b_a\frac{\omega}{n_B}L_{1_a})+\sum_a q_a^2L_{2_a}\sum_aq_a(L_{4_a}-b_a\frac{\omega}{n_B}L_{2_a})}{T\bigg[\bigg(\sum_a q_a^2L_{1_a}\bigg)^2+\bigg(\sum_a q_a^2L_{2_a}\bigg)^2\bigg]}\nonumber\\
 & = \frac{(\sigma_{el}/T)(\mathcal{I}_{31}^B/T^2)+(\sigma_{H}/T)(\mathcal{I}_{42}^B/T^2)}{(\sigma_{el}/T)^2+(\sigma_{H}/T)^2},
 \label{equnew49}
\end{align}
and the dimensionless Nernst coefficient (Nernst coefficient($N$) times the magnitude of the magnetic field ($B$)) is given as,
\begin{align}
 NB & = \frac{\sum_a q_a^2L_{1_a}\sum_aq_a(L_{4_a}-b_a\frac{\omega}{n_B}L_{2_a})-\sum_a q_a^2L_{2_a}\sum_aq_a(L_{3_a}-b_a\frac{\omega}{n_B}L_{1_a})}{T\bigg[\bigg(\sum_a q_a^2L_{1_a}\bigg)^2+\bigg(\sum_a q_a^2L_{2_a}\bigg)^2\bigg]}\nonumber\\
 & = \frac{(\sigma_{el}/T)(\mathcal{I}_{42}^B/T^2)-(\sigma_{H}/T)(\mathcal{I}_{31}^B/T^2)}{(\sigma_{el}/T)^2+(\sigma_{H}/T)^2}.
 \label{equnew50}
 \end{align}
Here we have identified the electrical conductivity
in the presence of a magnetic field and the Hall conductivity as, $\sigma_{el}=\sum_aq_a^2L_{1_a}$ and $\sigma_{H}=\sum_aq_a^2L_{2_a}$ respectively \cite{ranjitaarpanhm1}. The integrals $\mathcal{I}_{31}^B$ and $\mathcal{I}_{42}^B$ in Eqs.\eqref{equnew49} and \eqref{equnew50} are defined as  $\mathcal{I}_{31}^B = \sum_aq_a(L_{3_a}-b_a\frac{\omega}{n_B}L_{1_a})$ and $\mathcal{I}_{42}^B\equiv \sum_aq_a(L_{4_a}-b_a\frac{\omega}{n_B}L_{2_a})$.
Note that in the absence of a magnetic field integrals $L_{2_a}$ and $L_{4_a}$ are identically zero. Hence normalized Nernst coefficient ($NB$) vanishes in the absence of magnetic field and the magneto-Seebeck coefficient ($S_B$) boils down to the Seebeck coefficient ($S$) in the absence of magnetic field as given in Eq.\eqref{equnew11}. However it is important to note that unlike Eq.\eqref{equnew11}, in Eq.\eqref{equnew49} mesonic contributions do not cancel out in the numerator. In the numerator of Eq.\eqref{equnew49} mesonic contributions cancel out in the term $\mathcal{I}_{31}^B$ due to opposite contributions of particles and antiparticles. But in the term $\mathcal{I}_{42}^B$ mesonic contributions do not cancel out because mesons and its anti particles contribute equally.
Similarly the particles and antiparticles contribute in a constructive way in the Nernst coefficient. For a single baryon species the magneto-Seebeck coefficient and the normalized Nernst coefficient  can be expressed as,
\begin{align}
 S_{B_a} = \frac{1}{q_aT}\bigg(\frac{L_{1_a}L_{3_a}+L_{2_a}L_{4_a}}{L_{1_a}^2+L_{2_a}^2}-b_a\frac{\omega}{n_B}\bigg),
\label{equnew51}
 \end{align}
and,
\begin{align}
 N_aB = \frac{1}{q_aT}\bigg(\frac{L_{1_a}L_{4_a}-L_{2_a}L_{3_a}}{L_{1_a}^2+L_{2_a}^2}\bigg),
 \label{equnew52}
\end{align}
respectively. Note that the magneto-Seebeck coefficient and the normalized Nernst coefficient for a single baryon species as given in Eq.\eqref{equnew51} and Eq.\eqref{equnew52} are very similar to the magneto-Seebeck coefficient and the Nernst coefficient for condensed matter systems \cite{magSeeb1}, apart from the fact that in the condensed matter system $\omega/n_B$ factor does not appear rather chemical potential appears in the expression of magneto-Seebeck coefficient. One can also define Peltier coefficient and thermal conductivity in the presence of thermoelectric effect for nonvanishing magnetic field. But due to the presence of magnetic field these expressions become more complicated and we have not discussed it further. Using Eq.\eqref{equnew49} and Eq.\eqref{equnew50} we can estimate magneto-Seebeck coefficient and normalized Nernst coefficient for hadron resonance gas model by evaluating the integrals $L_{1_a}$, $L_{2_a}$, $L_{3_a}$ and $L_{4_a}$ as given in Eqs.\eqref{equnew40},\eqref{equnew41},\eqref{equnew42} and Eq.\eqref{equnew43} 
respectively. Note that to evaluate these integrals in HRG model we use Boltzmann approximation. In the Boltzmann approximation we get,  
\begin{align}
 & L_{1_a}=\frac{g_a}{3T}\int\frac{d^3p_a}{(2\pi)^3}\frac{\tau_a}{1+(\omega_{c_a}\tau_a)^2}\left(\frac{\vec{p}_a^2}{\epsilon_a^2}\right)f_a^{(0)},\label{equnew53}\\
& L_{2_a}=\frac{g_a}{3T}\int\frac{d^3p_a}{(2\pi)^3}\frac{\tau_a(\omega_{c_a}\tau_a)}{1+(\omega_{c_a}\tau_a)^2}\left(\frac{\vec{p}_a^2}{\epsilon_a^2}\right) f_a^{(0)},\label{equnew54}\\
& L_{3_a}=\frac{g_a}{3T}\int\frac{d^3p_a}{(2\pi)^3}\frac{\tau_a\epsilon_a}{1+(\omega_{c_a}\tau_a)^2}\left(\frac{\vec{p}_a^2}{\epsilon_a^2}\right)f_a^{(0)},\label{equnew55}\\
& L_{4_a}=\frac{g_a}{3T}\int\frac{d^3p_a}{(2\pi)^3}\frac{\tau_a\epsilon_a(\omega_{c_a}\tau_a)}{1+(\omega_{c_a}\tau_a)^2}\left(\frac{\vec{p}_a^2}{\epsilon_a^2}\right)f_a^{(0)}.\label{equnew56}
 \end{align}
The only unknown quantity in the Eqs.\eqref{equnew53}-\eqref{equnew56} is the relaxation time. In general relaxation time depends on the energy and the momentum of the particles. However for simplicity one takes thermal averaged relaxation time by integrating energy dependent relaxation time over equilibrium distribution functions. We discuss the thermal averaged relaxation time for the HRG model in the next section.

\section{HADRON RESONANCE GAS MODEL}
\label{HRGmodel}
The fundamental quantity to study the thermodynamics of the HRG model is the partition function at finite temperature ($T$) and baryon chemical potential ($\mu_B$), which can be written as \cite{KadamHM2015},
\begin{equation}
 \log Z(\beta,\mu_B,V)=\int dm (\rho_M(m)\log Z_b(m,V,\beta,\mu_B)+\rho_B(m)\log Z_f(m,V,\beta,\mu_B)),
 \label{equnew57}
\end{equation}
here, $V$ denotes the volume and $\beta = 1/T$ denotes the inverse of the temperature. Further the total partition function is composed of the partition functions of free mesons ($Z_b$) and baryons ($Z_f$) with mass $m$.  
Moreover, $\rho_B$ and $\rho_m$ denotes the spectral functions of free mesons and baryons respectively. The spectral function represents number of states within the mass range $m$ and $m+dm$. In this investigation we have considered all the  hadrons and their resonances below
a certain mass cutoff $\Lambda$, by taking the spectral density
$\rho_{B/M}(m)$ in the following form, 
\begin{equation}
 \rho_{B/M}(m)=\sum_i^{M_i<\Lambda}g_i\delta(m-m_i),
 \label{equnew58}
\end{equation}
 In  Eq.\eqref{equnew58}, $m_i$ and $g_i$ are
the masses and the  corresponding degeneracy of the known hadrons and their resonances respectively. It has been argued that along with the discrete
particle spectrum if one also includes
Hagedron spectrum then lattice QCD data for QCD trace anomaly can be reproduced up to $T \sim$ 160 MeV \cite{HRGMuller}. Otherwise with only discrete particle spectrum one can explain
lattice QCD data for trace anomaly only up to $T \sim 130 $ MeV \cite{HRGMuller}. One can also take into account the finite size effect of the hadrons into the calculation which has been discussed in the excluded volume HRG model. For details of thermodynamics 
of HRG model and various improvement, see e.g. Ref.\cite{HRG1} and references therein. In this investigation we do not consider any excluded volume effect, however only for the calculation of the scattering cross-section assuming hard sphere scattering we take a finite size of the hadrons. 

The relaxation time of particle $a$ having momentum $p_a$ and energy $\epsilon_a$ is defined as \cite{lataguruhm,KapustaChakraborty2011},
\begin{align}
\tau_a^{-1}(\epsilon_a)=\sum_{b,c,d}\int\frac{d^3p_b}{(2\pi)^3}\frac{d^3p_c}{(2\pi)^3}\frac{d^3p_d}{(2\pi)^3}W(a,b\rightarrow c,d)f_b^{(0)},
\end{align}
where the transition rate $W(a,b\rightarrow c,d)$ can be expressed in terms of the transition amplitude $\mathcal{M}$ in the following manner,

\begin{align}
W(a,b\rightarrow c,d)=\frac{(2\pi)^4\delta(p_a+p_b-p_c-p_d)}{2E_a2E_b2E_c2E_d}|\mathcal{M}|^2.
\end{align}

In the center of mass (COM) frame, the relaxation time ($\tau_a$) or equivalently interaction frequency ($\omega_a$) can be written as, 

\begin{align}
\tau_a^{-1}(\epsilon_a)\equiv \omega_a(\epsilon_a)=\sum_b\int\frac{d^3p_b}{(2\pi)^3}\sigma_{ab}v_{ab}f_b^{(0)}.
\label{equnew61}
\end{align}
In Eq.\eqref{equnew61}, $\sigma_{ab}$ denotes the total scattering cross section for the process, $a(p_a)+b(p_b)\rightarrow c(p_c)+d(p_d)$ and $v_{ab}$ is the relativistic relative velocity between the particles $a$ and $b$ which can be expressed as,

\begin{align}
 v_{ab}=\frac{\sqrt{(p_a.p_b)^2-m_a^2m_b^2}}{\epsilon_a\epsilon_b}.
\end{align}

In Eq.\eqref{equnew61} relaxation time is energy dependent. For simplicity one can consider energy averaged relaxation time which is by definition energy independent. The energy independent relaxation time $\tau^a$ can be obtained by averaging the relaxation time $\tau^a(\epsilon_a)$ over the distribution function $f_a^{(0)}(\epsilon_a)$ \cite{KapustaChakraborty2011,paramitahm2016},

\begin{align}
 \tau_a^{~-1}=\frac{\int f_a^{(0)}\tau^{-1}_a(\epsilon_a)d\epsilon_a}{\int f^{(0)}_ad\epsilon_a}
 \label{equnew63}
\end{align}

Eq.\eqref{equnew61} and Eq.\eqref{equnew63} allows us to write the energy averaged relaxation time
$(\tau_a)$ in terms of the scattering cross section in the following manner \cite{GURUHM2015},

\begin{equation}
 \tau_a^{~-1}=\sum_b n_b\langle\sigma_{ab}v_{ab}\rangle,
 \label{equnew64}
\end{equation}
here $n_b$ is the number density of particle ``b'' and $\langle\sigma_{ab}v_{ab}\rangle$ represents thermal averaged cross section. The thermal averaged 
cross section for the scattering process $a(p_a)+b(p_b)\rightarrow c(p_c)+d(p_d)$ can be shown to be \cite{gondologelmini},

\begin{align}
 \langle\sigma_{ab}v_{ab}\rangle = \frac{\int d^3p_ad^3p_b \sigma_{ab}v_{ab}f_a^{(0)}(p_a)f_a^{(0)}(p_b)}{\int d^3p_ad^3p_b f_a^{(0)}(p_a)f_a^{(0)}(p_b)}.
\end{align}
Under Boltzmann approximation with the equilibrium distribution function of the form $f_a^{(0)}(p_a) = e^{-(\epsilon_a\pm\mu_a)/T}$, the thermal averaged cross section for the scattering of hard spheres (of radius $r_h$) becomes, 
\begin{align}
 \langle\sigma_{ab}v_{ab}\rangle = \frac{\sigma\int d^3p_ad^3p_b v_{ab}e^{-\epsilon_a/T}e^{-\epsilon_b/T}}{\int d^3p_ad^3p_b e^{-\epsilon_a/T}e^{-\epsilon_b/T}}.
\label{equnew66}
 \end{align}

Note that in the Boltzmann approximation chemical potential dependence gets canceled from the numerator and denominator in Eq.\eqref{equnew66}. One can change the momentum integration to the center of mass energy variable ($\sqrt{s}$) to get, 

\begin{align}
 \int d^3p_ad^3p_b v_{ab}e^{-E_a/T}e^{-E_b/T} = 2\pi^2T\int ds\sqrt{s}(s-4m^2)K_1(\sqrt{s}/T),
\end{align}
and
\begin{align}
 \int d^3p_ad^3p_b e^{-E_a/T}e^{-E_b/T} = \left(4\pi m^2 T K_2(m/T)\right)^2.
\end{align}

Thus the thermal averaged cross section as given in Eq.\eqref{equnew66} can be written as, 

\begin{align}
 \langle\sigma_{ab}v_{ab}\rangle =\frac{\sigma}{8m^4TK_2^2(m/T)}\int_{4m^2}^{\infty}ds \sqrt{s}(s-4m^2)K_1(\sqrt s /T).
 \label{equnew69}
\end{align}

Here $K_1$ and $K_2$ are modified Bessel function of first order and second order respectively. Thermal averaged cross section as given in Eq.\eqref{equnew69}can be generalized for the scattering of different species. In this case the the thermal averaged scattering cross section can be expressed as,

\begin{equation}
 \langle\sigma_{ab}v_{ab}\rangle = \frac{\sigma}{8Tm_a^2m_b^2K_2(m_a/T)K_2(m_b/T)}\int_{(m_a+m_b)^2}^{\infty}ds\times \frac{[s-(m_a-m_b)^2]}{\sqrt{s}}
 \times [s-(m_a+m_b)^2]K_1(\sqrt{s}/T),
 \label{equnew70}
\end{equation}
here $\sigma = 4\pi r_h^2$ is the total scattering cross section for the hard spheres. Using Eq.\eqref{equnew70} and Eq.\eqref{equnew64} one can calculate the thermal averaged relaxation time. It is important to mention that while the hard sphere scattering cross section $\sigma$ is independent of both temperature and baryon chemical potential, thermal averaged cross section $\langle\sigma v\rangle$ can depend on temperature ($T$) and chemical potential ($\mu_B$). This temperature ($T$) and chemical potential ($\mu_B$) dependence arise due to the temperature ($T$) and chemical potential ($\mu_B$) dependence of the distribution functions. 
 However in Boltzmann approximation $\langle\sigma v\rangle$ is independent of $\mu_B$  \cite{gondologelmini}. After evaluating the thermal averaged relaxation time using Eq. \eqref{equnew64} for each hadrons we can estimate the magneto-Seebeck coefficient and Nernst coefficient using Eq.\eqref{equnew49} and \eqref{equnew50} respectively.

 \begin{figure}[]
    \centering
        \includegraphics[width=0.65\linewidth]{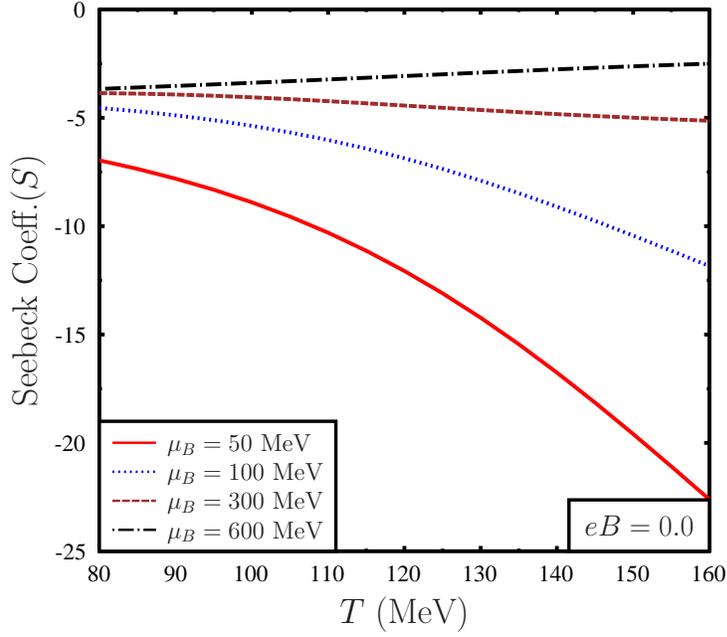}
   \caption{ Variation of Seebeck coefficient ($S$) for vanishing magnetic field with temperature ($T$) and baryon chemical potential ($\mu_B$). With increasing baryon chemical potential ($\mu_B$), Seebeck coefficient ($S$) increases. On the other hand with increasing temperature Seebeck coefficient decreases for values of $\mu_B$ ($\lesssim 300$ MeV), but for higher values of baryon chemical potential Seebeck coefficient increases with temperature.}
\label{fig1}
\end{figure}

\begin{figure}
    \centering
    \begin{minipage}{.48\textwidth}
        \centering
        \includegraphics[width=1.2\linewidth]{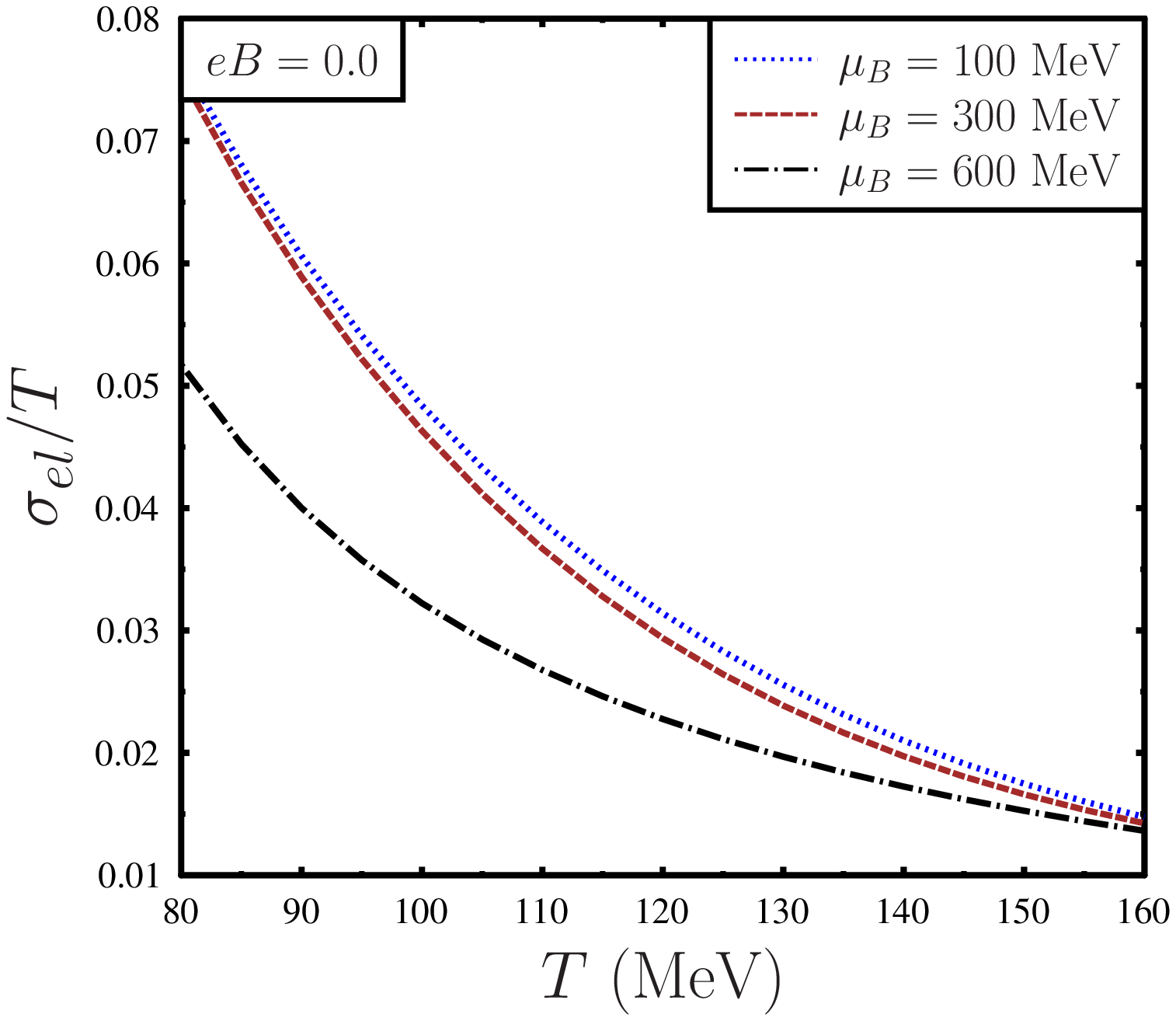}
    \end{minipage}~~
    \begin{minipage}{0.48\textwidth}
        \centering
        \includegraphics[width=1.2\linewidth]{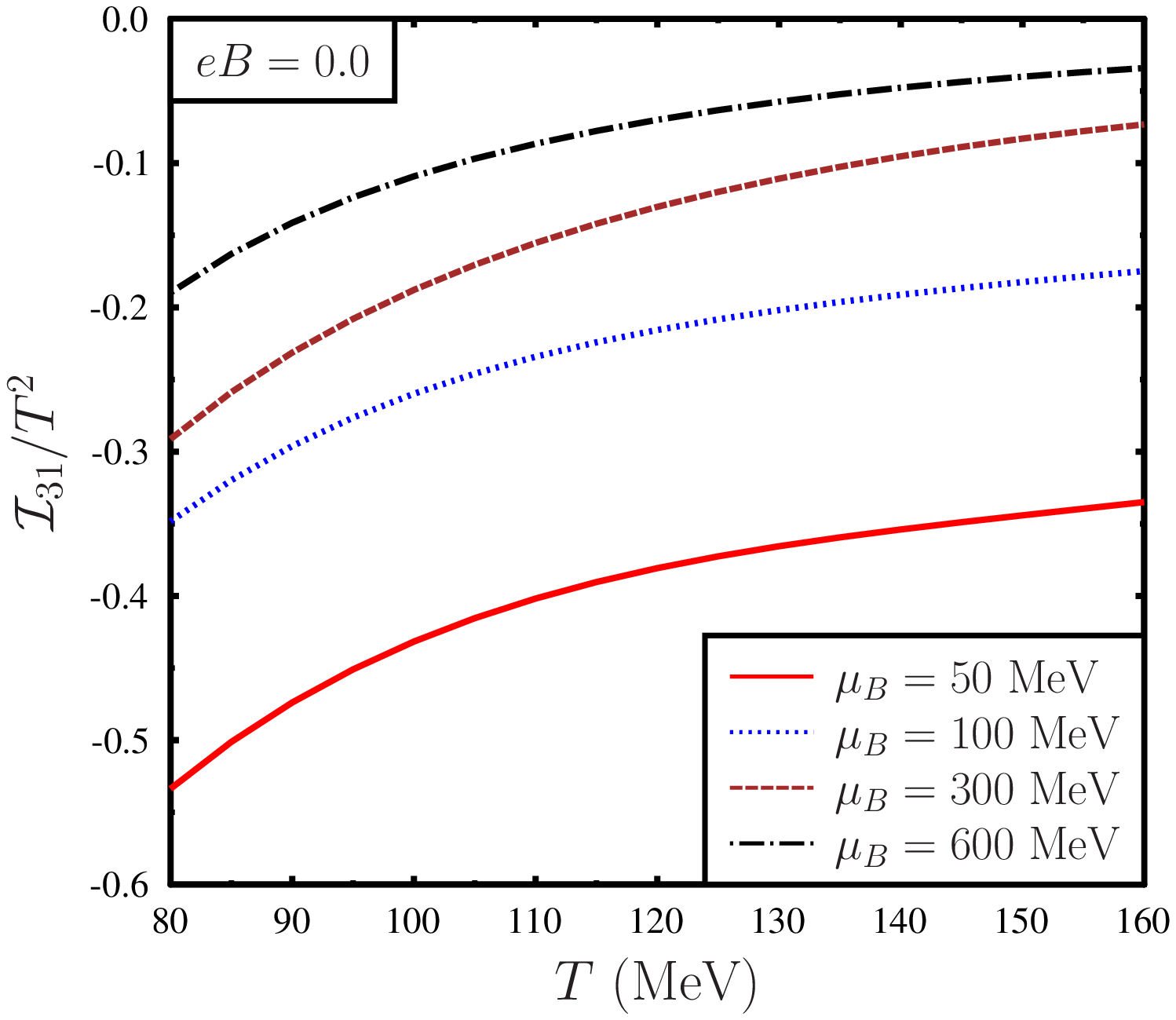}
    \end{minipage}
    \caption{Left plot: variation of normalised electrical conductivity ($\sigma_{el}/T$) with temperature ($T$) and baryon chemical potential ($\mu_B$) for vanishing magnetic field. With increasing $T$ and $\mu_B$ normalised electrical conductivity decreases. Right plot: variation of $\mathcal{I}_{31}/T^2$ with temperature and baryon chemical potential for vanishing magnetic field. With increasing $T$ and $\mu_B$, $\mathcal{I}_{31}/T^2$ increases.}
\label{fig1a}
 \centering
            \includegraphics[width=0.65\linewidth]{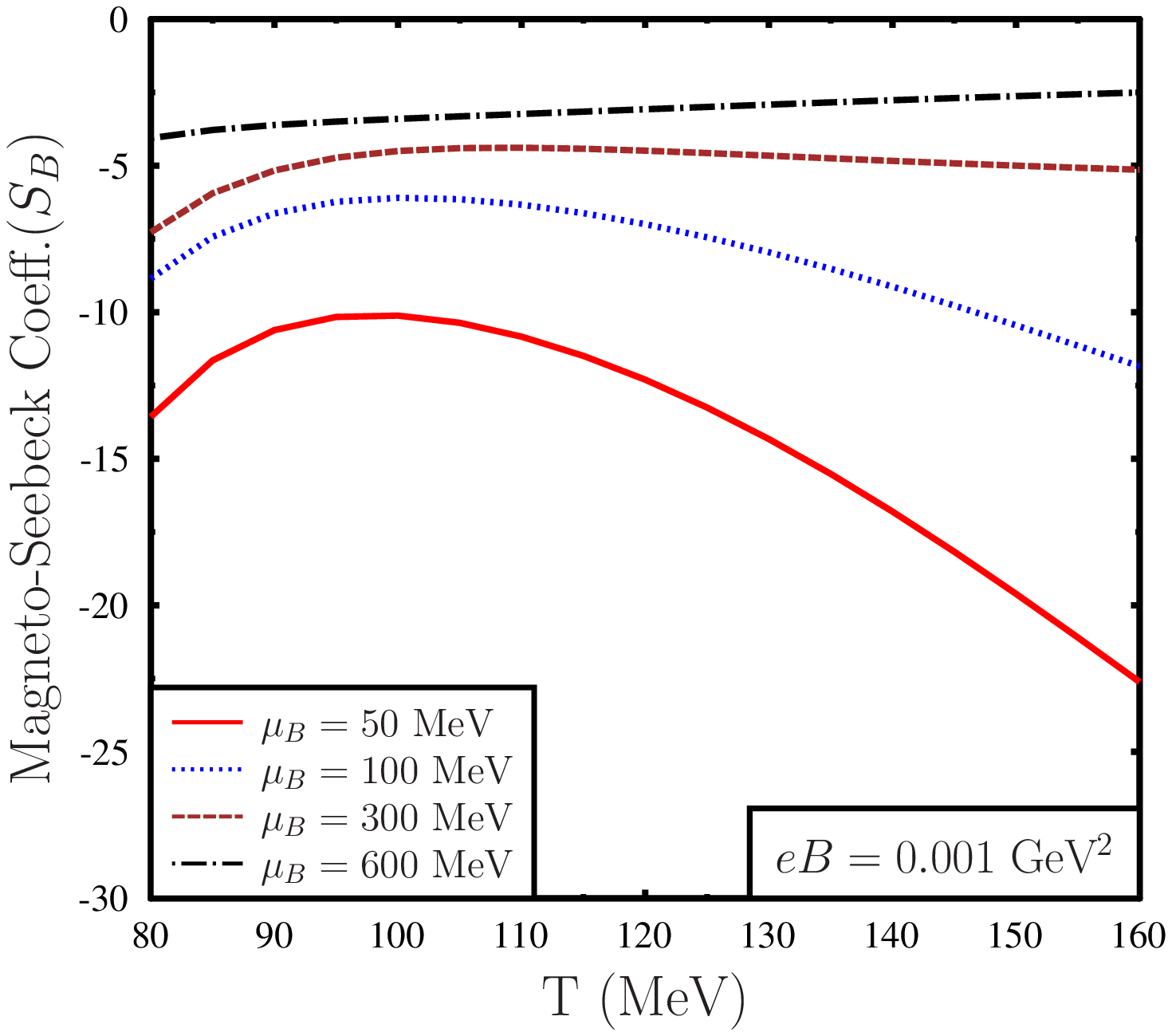}
    \caption{Variation of the magneto-Seebeck coefficient ($S_B$) with temperature ($T$) for different values of baryon chemical potential ($\mu_B$).  With increasing baryon chemical potential ($\mu_B$) the magneto-Seebeck coefficient ($S_B$) increases. On the other hand with increasing temperature the magneto-Seebeck coefficient shows nonmonotonic behavior for $\mu_B\lesssim300$ MeV and for higher values of $\mu_B$ the magneto-Seebeck coefficient increases.}
\label{fig2}
\end{figure}

\section{results and discussions}
\label{results}
As we have mentioned earlier, for the hadron resonance gas model with a discrete particle spectrum, we consider here all the hadrons and their resonances up to a mass cutoff $\Lambda =  2.6$ GeV as is listed in Ref.\cite{pdg}. A detailed list of particles has been given in Appendix A of Ref.\cite{kapustaAlbr}.
This apart relaxation time also enters into the calculation of thermoelectric coefficients. For the estimation of relaxation time within the approximation of hard-sphere scattering, we consider a uniform radius of $r_h=0.3$ fm for all the mesons and baryons 
\cite{GURUHM2015,hmgururadius1}. For these parameters, we estimate the magneto-Seebeck coefficient and the Nernst coefficient using Eq.\eqref{equnew49} and Eq.\eqref{equnew50} as a function of temperature $(T)$ and baryon chemical potential ($\mu_B$) for nonvanishing values of the magnetic field.

\subsection{Behavior of Seebeck coefficient with vanishing magnetic field}

In Fig.\eqref{fig1} we show the variation of Seebeck coefficient (S) for vanishing magnetic field with temperature ($T$) and baryon chemical potential ($\mu_B$). From fig.\eqref{fig1} it is clear that for the range of temperature and baryon chemical potential considered in this investigation Seebeck coefficient (S) of the hot and dense hadron resonance gas is negative for vanishing magnetic field. This result is in contrast with our previous work \cite{arpanhm}. In Ref. \cite{arpanhm} Seebeck coefficient for vanishing magnetic field was found to be positive. The reason behind this discrepancy between the results obtained here and with the results in Ref. \cite{arpanhm} is that the formalism adopted in this investigation is different from that of Ref. \cite{arpanhm}. For vanishing magnetic field the Seebeck coefficient as given in Eq.\eqref{equnew11} is different with respect to the expression of the Seebeck coefficient as obtained in Ref. \cite{arpanhm}. In Eq.\eqref{equnew11} we have a factor of $(\epsilon_a-b_a\omega/n_B)$, which is different in Ref.\cite{arpanhm} and was $(\epsilon_a-b_a\mu_B)$. For relativistic systems, heat flow can only be defined relative to a conserved current, for strongly interacting plasma, the heat current is defined with respect to the net baryon current. Let us note that $\omega/n_B = Ts/n_B+\mu_B$. For HRG model the entropy arising for pions is large making $Ts/n_B>>\mu_B$ and in fact overwhelms single particle energy $\epsilon_a$. This makes ($\epsilon-\omega/n_B$) negative for baryons in the HRG model. 
Baryons have dominant contributions in the Seebeck coefficient for vanishing magnetic field. Hence for the HRG model the factor $(\epsilon-(\omega/{n_B}))$ makes the total Seebeck coefficient ($S$) negative. At this point, it is perhaps relevant to note that the other transport coefficient thermal conductivity ($k_0$) in the absence of any thermoelectric effect as defined in Eq.\eqref{equnew15} also has the term $(\epsilon-(\omega/{n_B}))$. However, in the expression of thermal conductivity $(\epsilon-(\omega/{n_B}))$ always comes as a square, hence in the absence of thermoelectric effect, the positivity of thermal conductivity is ensured which can be seen in Eq.\eqref{equnew15}. It may be noted that both positive and negative values of the Seebeck coefficient can be found in condensed matter systems, e.g., if for the holes the Seebeck coefficient is positive then for electrons the Seebeck coefficient is negative. 

\begin{figure}[]
    \centering
    \begin{minipage}{.48\textwidth}
        \centering
        \includegraphics[width=1.2\linewidth]{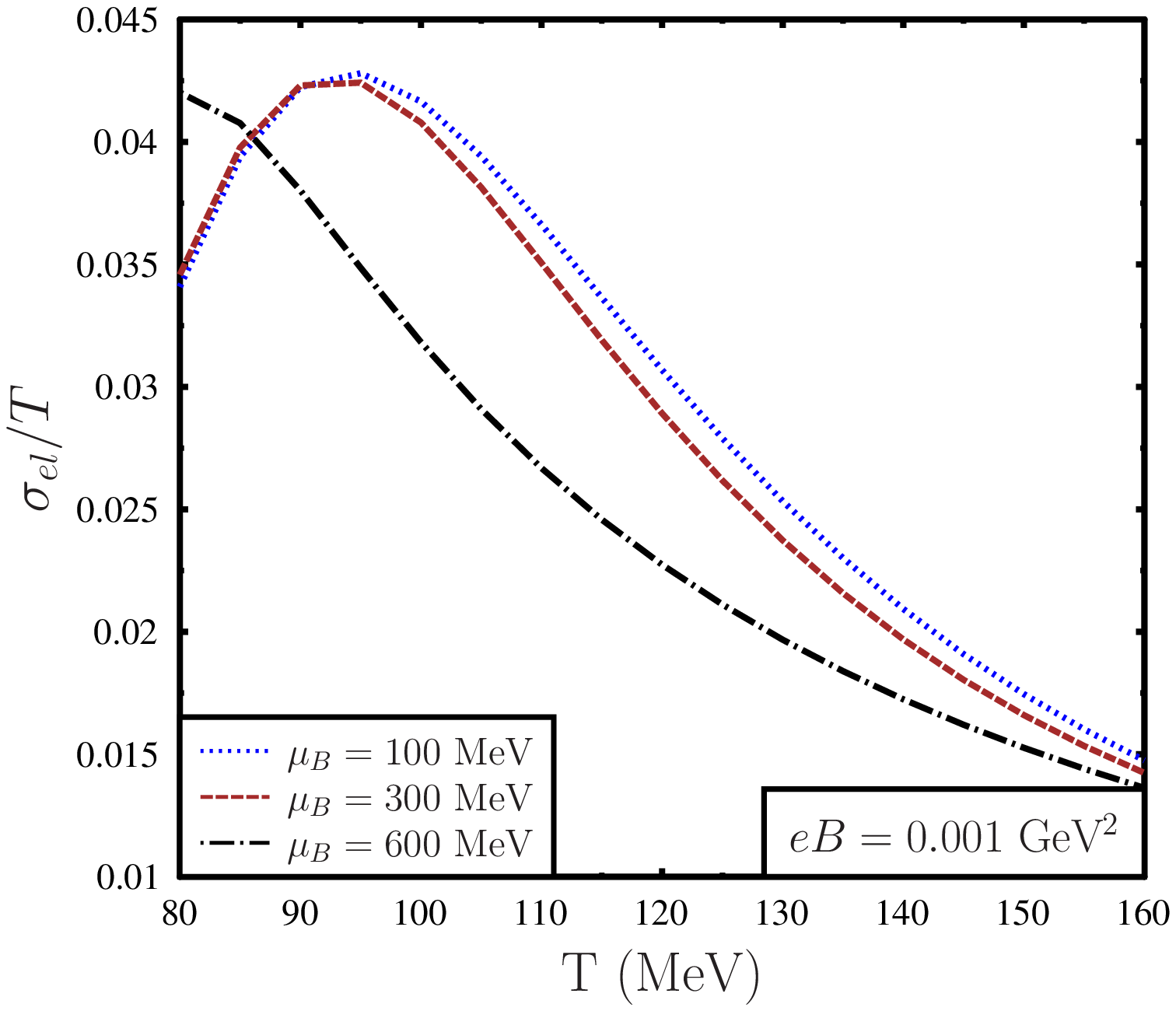}
    \end{minipage}~~
    \begin{minipage}{0.48\textwidth}
        \centering
        \includegraphics[width=1.2\linewidth]{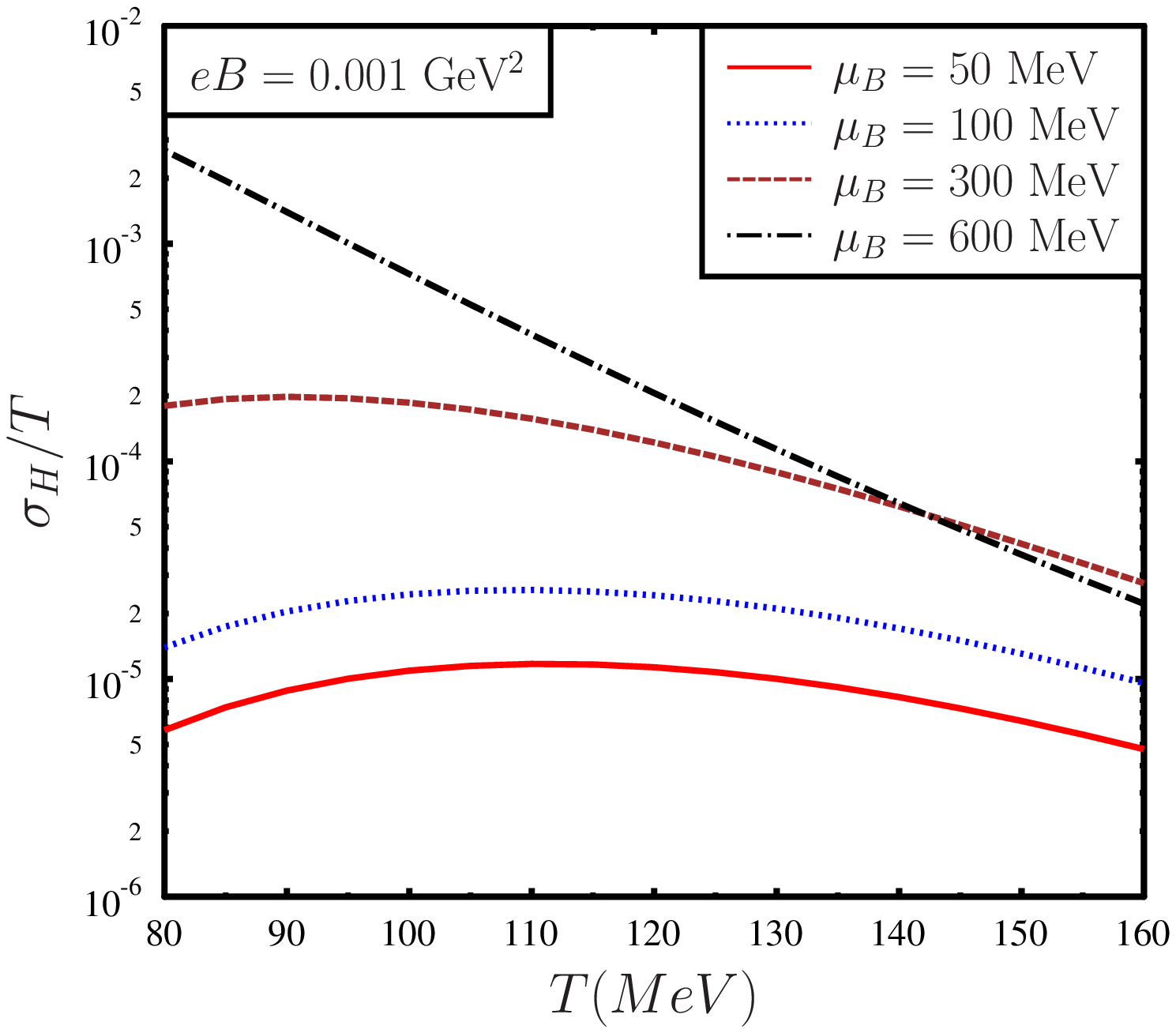}
    \end{minipage}
    \caption{Left plot: variation of normalized electrical conductivity $\sigma_{el}/T$ with temperature for different values of baryon chemical potential for nonvanishing magnetic field. For low temperature range $\sigma_{el}/T$ increases with $\mu_B$, otherwise it decreases with baryon chemical potential. With temperature $\sigma_{el}/T$ shows nonmonotonic variation with a peak structure. Right plot: variation of normalized Hall conductivity $\sigma_{H}/T$ with temperature for different values of baryon chemical potential.}
\label{fig3}
    \centering
    \begin{minipage}{.48\textwidth}
        \centering
        \includegraphics[width=1.2\linewidth]{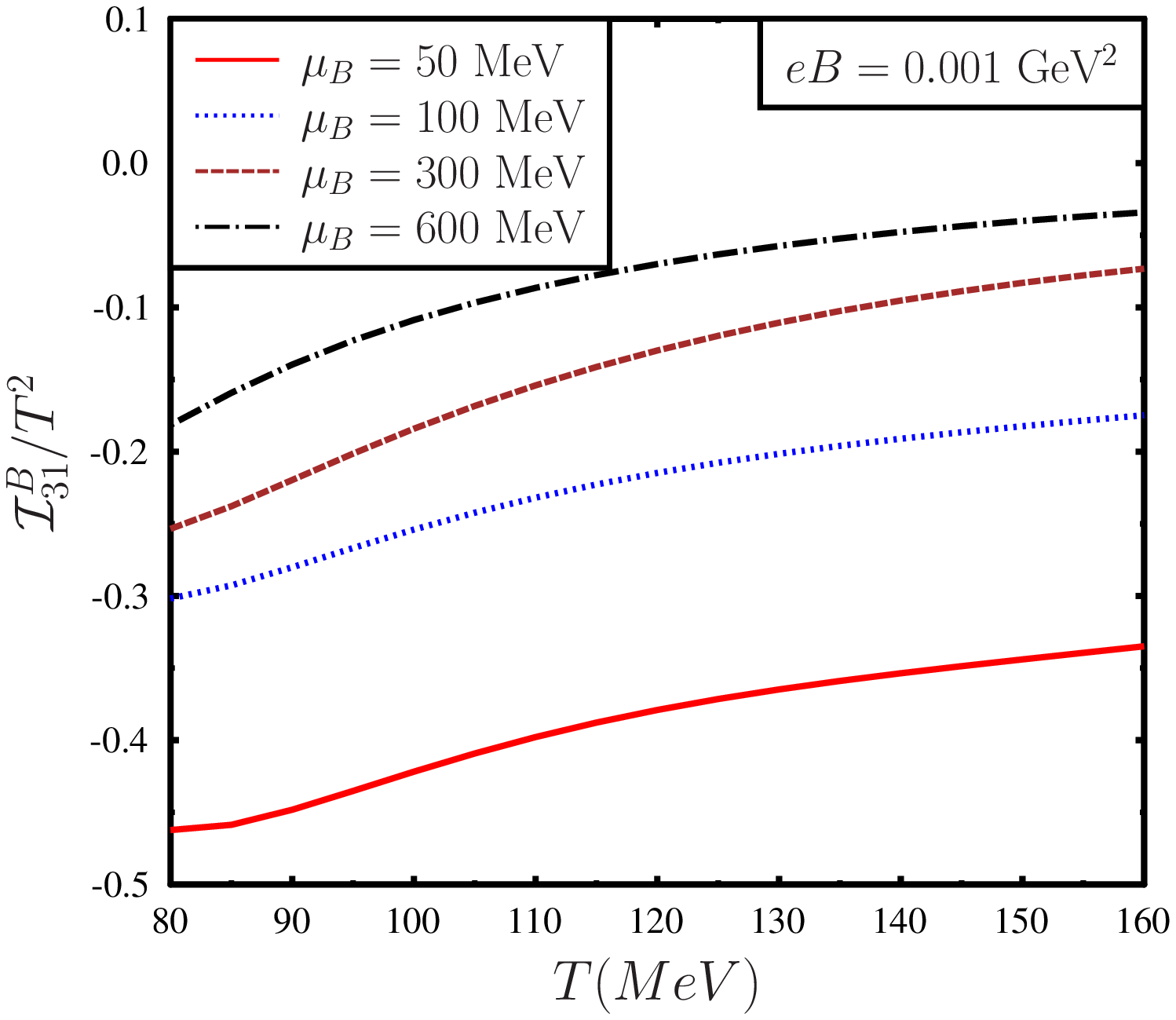}
    \end{minipage}~~
    \begin{minipage}{0.48\textwidth}
        \centering
        \includegraphics[width=1.2\linewidth]{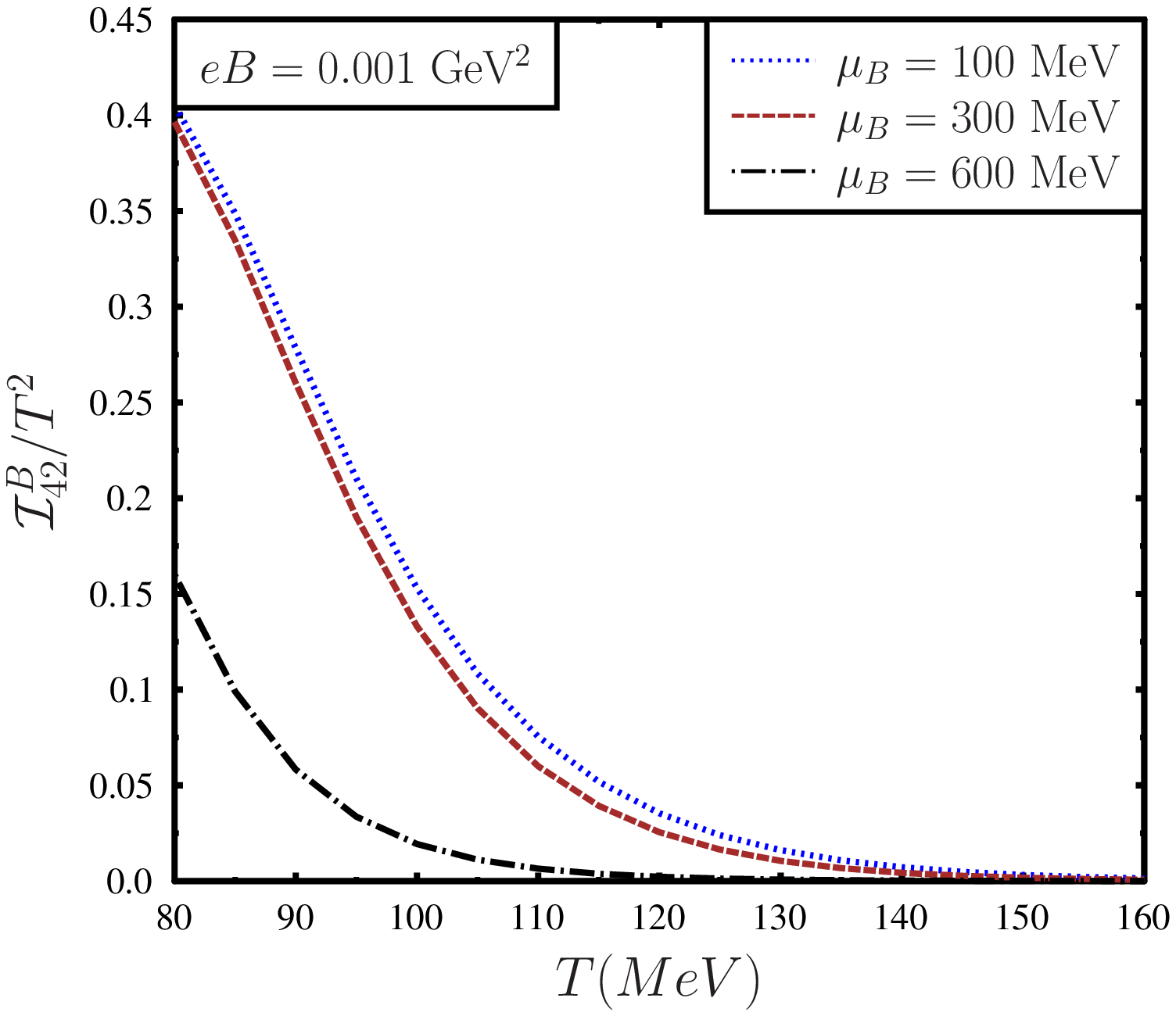}
    \end{minipage}
    \caption{Left plot: variation of $\mathcal{I}_{31}^B/T^2$ with temperature for different values of baryon chemical potential for nonvanishing magnetic field. With temperature as well as with baryon chemical potential $\mathcal{I}_{31}^B/T^2$ increase. Right plot: variation of $\mathcal{I}_{42}^B/T^2$ with temperature for different values of baryon chemical potential for nonvanishing magnetic field. With increasing temperature as well as with $\mu_B$, $\mathcal{I}_{42}^B/T^2$ decreases. In $\mathcal{I}_{31}^B/T^2$ mesonic contributions vanishes due to exact and opposite contributions of the particles and the antiparticles. On the other hand in $\mathcal{I}_{42}^B/T^2$ mesonic contributions are dominant relative to the baryonic contributions.}
\label{fig4}
\end{figure}

In Fig.\eqref{fig1} it may be observed that with an increase in temperature ($T$) the Seebeck coefficient ($S$) decreases for $\mu_B\lesssim 300$ MeV and it increases with temperature for higher values of $\mu_B$. On the other hand with the baryon chemical potential ($\mu_B$) the Seebeck coefficient ($S$) always increases. Let us recall that $S = (\mathcal{I}_{31}/T^2)/(\sigma_{el}/T)$ therefore variation of the Seebeck coefficient with temperature and baryon chemical potential for vanishing magnetic field can be understood by looking into the behaviour of $\mathcal{I}_{31}/T^2$ and $\sigma_{el}/T$ with $T$ and $\mu_B$ as shown in Fig.\eqref{fig1a}. Variation of $\sigma_{el}/T$ in the absence of magnetic field has been extensively discussed in Ref.\cite{ranjitaarpanhm1}. From the left plot in Fig.\eqref{fig1a} we can observe that for vanishing magnetic field normalized electrical conductivity ($\sigma_{el}/T$) decrease with temperature and baryon chemical potential. Among all the hadrons mesonic contribution to $\sigma_{el}/T$ is dominant relative to the baryonic contribution. With increasing temperature and baryon chemical potential mesonic contribution to $\sigma_{el}/T$ decreases. This decrease of normalized electrical conductivity with temperature and baryon chemical potential is predominantly due to the decrease of relaxation time of mesons with $T$ and $\mu_B$ (for a detailed discussion see Ref.\cite{ranjitaarpanhm1}). Further from the right plot in Fig.\eqref{fig1a} we can see that $\mathcal{I}_{31}/T^2$ increases with temperature and baryon chemical potential. This increase of $\mathcal{I}_{31}/T^2$ is due to the increasing behavior of $(-\omega/n_B)$ and the equilibrium distribution function with $T$ and $\mu_B$ (for a detailed discussion on the variation of $\omega/n_B$ see Ref.\cite{ranjitaarpanhm2}). 
For smaller chemical potentials $\mu_B \lesssim 300$ MeV the decrease of $\sigma_{el}/T$ is faster compared to the increase of $\mathcal{I}_{31}/T^2$ which leads to an increase in magnitude of the ratio that is magnitude of $S$. For higher chemical potentials the increase of  $\mathcal{I}_{31}/T^2$ due to contributions of baryons becomes faster compared to the decrease of $\sigma_{el}/T$ leading to the decrease in the magnitude of the Seebeck coefficient.   
For $\mu_B\sim 300$ MeV and higher, the increase in $\mathcal{I}_{31}/T^2$ with increasing temperature 
and the decrease in normalized electrical conductivity conspire to give rise to an increase of the Seebeck coefficient with the temperature at high values of baryon chemical potential. Let us recall that in the expression of $\mathcal{I}_{31}$ as given in Eq.\eqref{I31equ} mesons do not contribute in the summation. This is because of the opposite electrical charge, the contributions of mesons and its antiparticles cancel exactly in $\mathcal{I}_{31}$. On the other hand in $\sigma_{el}$ mesonic contribution is nonvanishing. At finite baryon chemical potential baryon contributions are larger compared to the antibaryon contributions due to the equilibrium distribution function. Among all the baryons contribution of proton to $\mathcal{I}_{31}/T^2$ is dominant due to its less mass with respect to the other baryons.
We can also see from Fig.\eqref{fig1} that with increasing baryon chemical potential Seebeck coefficient increases. With increasing baryon chemical potential $\sigma_{el}/T$ decreases. Among various $\mu_B$ dependent factors in $\sigma_{el}/T$, distribution function increases with $\mu_B$ and the relaxation time decreases with $\mu_B$ due to large number density of the scatters. It turns out that in HRG model pions are the dominant contributors in the $\sigma_{el}/T$. Although the number density of pion does not depend on the baryon chemical potential, due to pion-nucleon scattering pion relaxation time decreases with $\mu_B$, due to which $\sigma_{el}/T$ decreases with $\mu_B$. On the other hand, $\mathcal{I}_{31}/T^2$ increases with $\mu_B$ as can be seen in Fig.\eqref{fig1a}. For the range of temperature and baryon chemical potential considered here this variation of $\mathcal{I}_{31}/T^2$ and $\sigma_{el}/T$ with $\mu_B$ results in an increasing behavior of the Seebeck coefficient ($S$) with $\mu_B$.

\subsection{Results for magneto-Seebeck coefficient and Nernst coefficient}

Now let us discuss the behavior of Seebeck coefficient in the presence of a nonvanishing magnetic field with temperature and baryon chemical potential. In Fig.\eqref{fig2} we show the variation of the magneto-Seebeck coefficient ($S_B$) with temperature ($T$) and baryon chemical potential ($\mu_B$). Magneto-Seebeck coefficient ($S_B$) as shown in Fig.\eqref{fig2} shows a nonmonotonic behavior with temperature for $\mu_B\lesssim 300$ MeV. On the other hand for higher values of $\mu_B$, magneto-Seebeck coefficient ($S_B$) increases with temperature. We can also observe that with $\mu_B$ magneto-Seebeck coefficient always increases. The expression of the magneto-Seebeck coefficient as given in Eq.\eqref{equnew49} is more complicated than its zero magnetic field counterpart. 
From Eq.\eqref{equnew49} we can see that in  terms $\mathcal{I}_{31}^B/T^2$ and $\sigma_H/T$, mesonic contribution exactly cancels due to the exact and opposite contribution coming from particles and its antiparticles. But in terms $\mathcal{I}_{42}^B/T^2$ and $\sigma_{el}/T$ mesonic contributions do not cancel out. In fact in $\mathcal{I}_{42}^B/T^2$ and $\sigma_{el}/T$ mesons are the dominant contributors relative to the baryons.  
The variation of $S_B$ with temperature and baryon chemical potential crucially depend upon the variation of
$\sigma_{el}/T$, $\sigma_{H}/T$, $\mathcal{I}_{31}^B/T^2$ and $\mathcal{I}_{42}^B/T^2$ 
with $T$ and $\mu_B$. We discuss the variation of these four quantities in what follows.

 In the presence of magnetic field the variations of normalized electrical conductivity $(\sigma_{el}/T)$ and 
the Hall conductivity ($\sigma_{H}/T$) has been extensively discussed in Ref.\cite{ranjitaarpanhm1}, but for relatively large values of magnetic field compared to this investigation. For completeness in Fig.\eqref{fig3} we show the variation of $\sigma_{el}/T$ and $\sigma_{H}/T$.  
 For nonvanishing magnetic field normalized electrical conductivity and the Hall conductivity shows nonmonotonic variation with temperature, basically due to the factors of $\omega_c\tau$ in the expressions of $L_{1_a}$ and $L_{2_a}$, as has been discussed in details in Ref.\cite{ranjitaarpanhm1}.
 From the left plot in Fig.\eqref{fig3} we can see that for low temperature range (below $T\sim 90$ MeV), normalized electrical conductivity increases with $\mu_B$, but at higher temperature range it decreases with $\mu_B$. In the normalized electrical conductivity there are two contributions, one is due to the mesons and other one is due to the baryons. Mesonic contribution to $\sigma_{el}/T$ decreases with $\mu_B$ on the other hand baryonic contribution increases with $\mu_B$. In the low temperature range when the relaxation time is large magnetic field affects the mesonic contribution significantly due to the large value of $\omega_c\tau$. But with increasing $\mu_B$ the baryonic contribution compensates the decrease in the mesonic contribution and eventually for large $\mu_B$, normalized electrical conductivity increases with $\mu_B$. However, in the high temperature
 range the decrease in the mesonic contribution cannot be compensated by the baryonic contribution. Thus in the high temperature range normalised electrical conductivity decreases with $\mu_B$. This apart normalised Hall conductivity generically increases with baryon chemical potential as has been discussed in details in Ref.\cite{ranjitaarpanhm1}. This increase in the Hall conductivity with $\mu_B$ is due to the increasing net baryonic contribution with $\mu_B$. 
 
 Next we discuss the $T,\mu_B$ dependence of the other two quantities $\mathcal{I}_{31}^B/T^2$ and $\mathcal{I}_{42}^B/T^2$ on which $S_B$ depends.
  In Fig.\eqref{fig4} it is observed that generically, $\mathcal{I}_{31}^B/T^2$  as well as  $\mathcal{I}_{42}^B/T^2$  are monotonic functions of temperature and baryon chemical potential at nonvanishing magnetic field.
  It is clear from the expression of $\mathcal{I}_{31}^B/T^2$ is that only baryons contribute to $\mathcal{I}_{31}^B/T^2$ and for baryonic contribution $\omega/n_B$ is very large with respect to single particle energy. With increasing temperature and baryon chemical potential the factor ($-\omega/n_B$) increases along with the increasing distribution function giving rise to increasing behavior of $\mathcal{I}_{31}^B/T^2$ with temperature and baryon chemical potential.
  On the other hand in $\mathcal{I}_{42}^B/T^2$ mesonic contributions becomes dominant with respect to the baryonic contribution. 
  For mesonic contributions, increase in the distribution function with temperature is compensated by the decrease in relaxation time with increasing temperature. Thus with temperature $\mathcal{I}_{42}^B/T^2$ decreases. 
  Further with increasing baryon chemical potential mesonic contribution decreases because of decrease in the relaxation time. 
    It is important to note that   
  the nonmonotonic nature of the magneto-Seebeck coefficient with temperature has its origin in the nonmonotonic variation of $\sigma_{el}/T$ and $\sigma_H/T$ with temperature. Further we can observe from Fig.\eqref{fig3} that $\sigma_{el}/T>> \sigma_{H}/T$ and the order of magnitude values of $\mathcal{I}_{31}^B/T^2$ and $\mathcal{I}_{42}^B/T^2$ are of the same order. Therefore the variation of magneto-Seebeck coefficient $S_B$ can be approximately expressed as, $S_B\sim (\mathcal{I}_{31}^B/T^2)/(\sigma_{el}/T)$. 
 Thus for small value of baryon chemical potential the maximum in Fig.\eqref{fig2} corresponds to the maximum of $\sigma_{el}/T$ as shown in Fig.\eqref{fig3} and at high temperature decreasing behavior of $S_B$ is due to the decrease of $\sigma_{el}/T$ with temperature. 
\begin{figure}[!htb]
    \centering
    \begin{minipage}{.48\textwidth}
        \centering
        \includegraphics[width=1.2\linewidth]{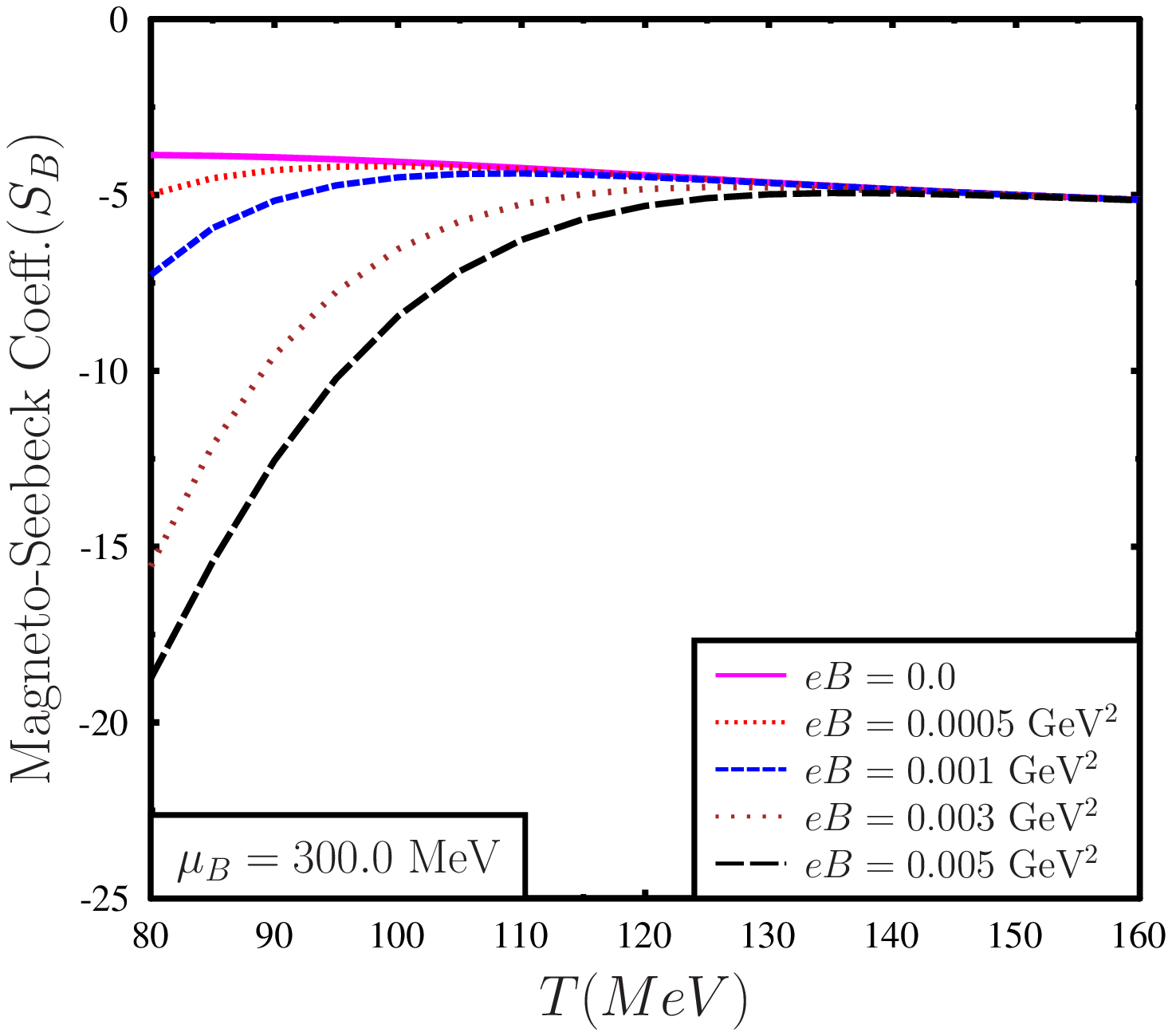}
    \end{minipage}~~
    \begin{minipage}{0.48\textwidth}
        \centering
        \includegraphics[width=1.2\linewidth]{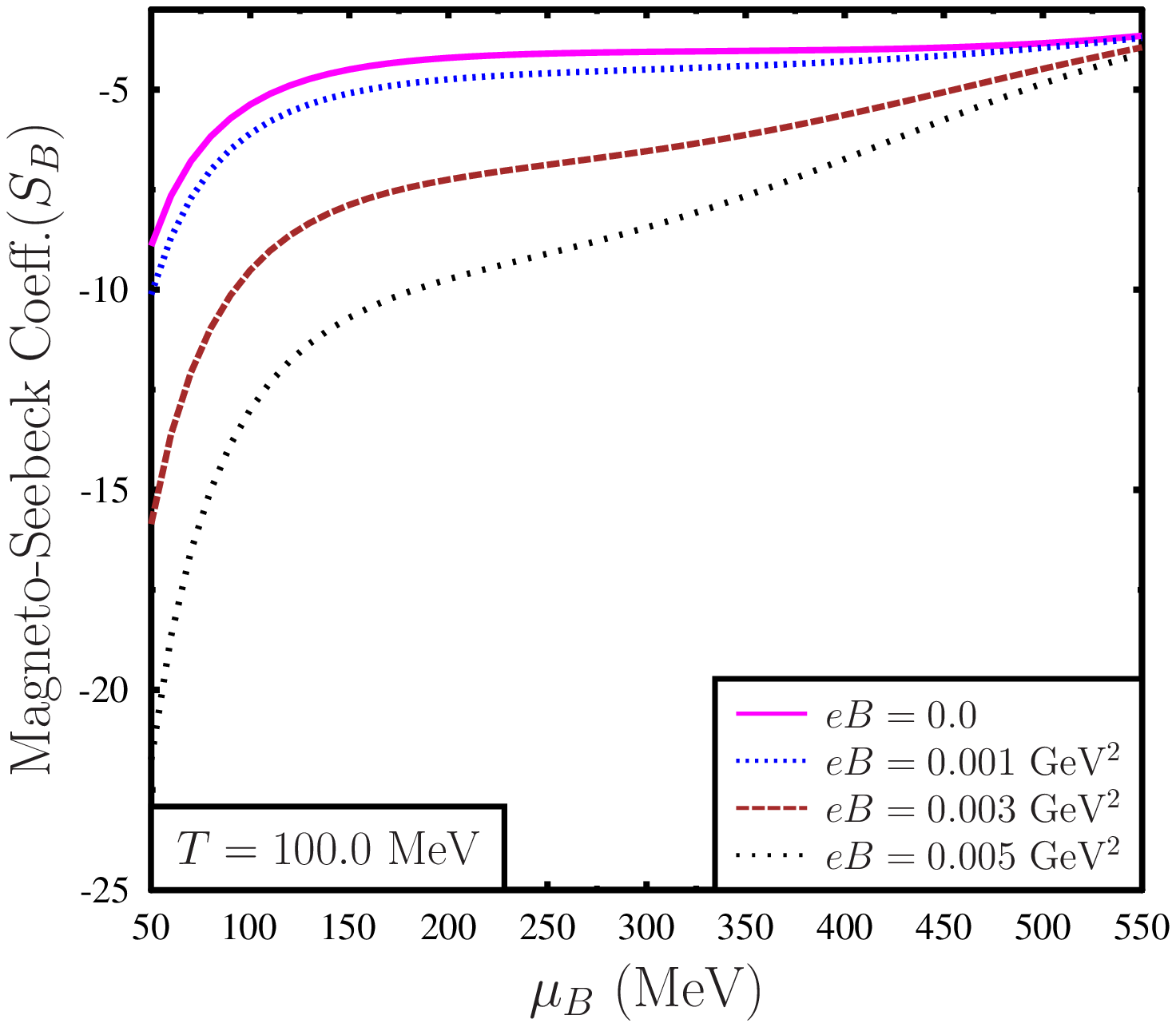}
    \end{minipage}
    \caption{Left plot: variation of magneto-Seebeck coefficient ($S_B$) with temperature for different values of magnetic field. With increasing magnetic field $S_B$ decreases, however in the high temperature range the effect of magnetic field on $S_B$ is less significant. Right plot:variation of magneto-Seebeck coefficient $S_B$ with baryon chemical potential for different values of magnetic field. From this plot it is clear that with increasing $\mu_B$ magneto-Seebeck coefficient ($S_B$) increases.}
\label{fig5}
\end{figure}

\begin{figure}[!htb]
    \centering
    \begin{minipage}{.48\textwidth}
        \centering
        \includegraphics[width=1.2\linewidth]{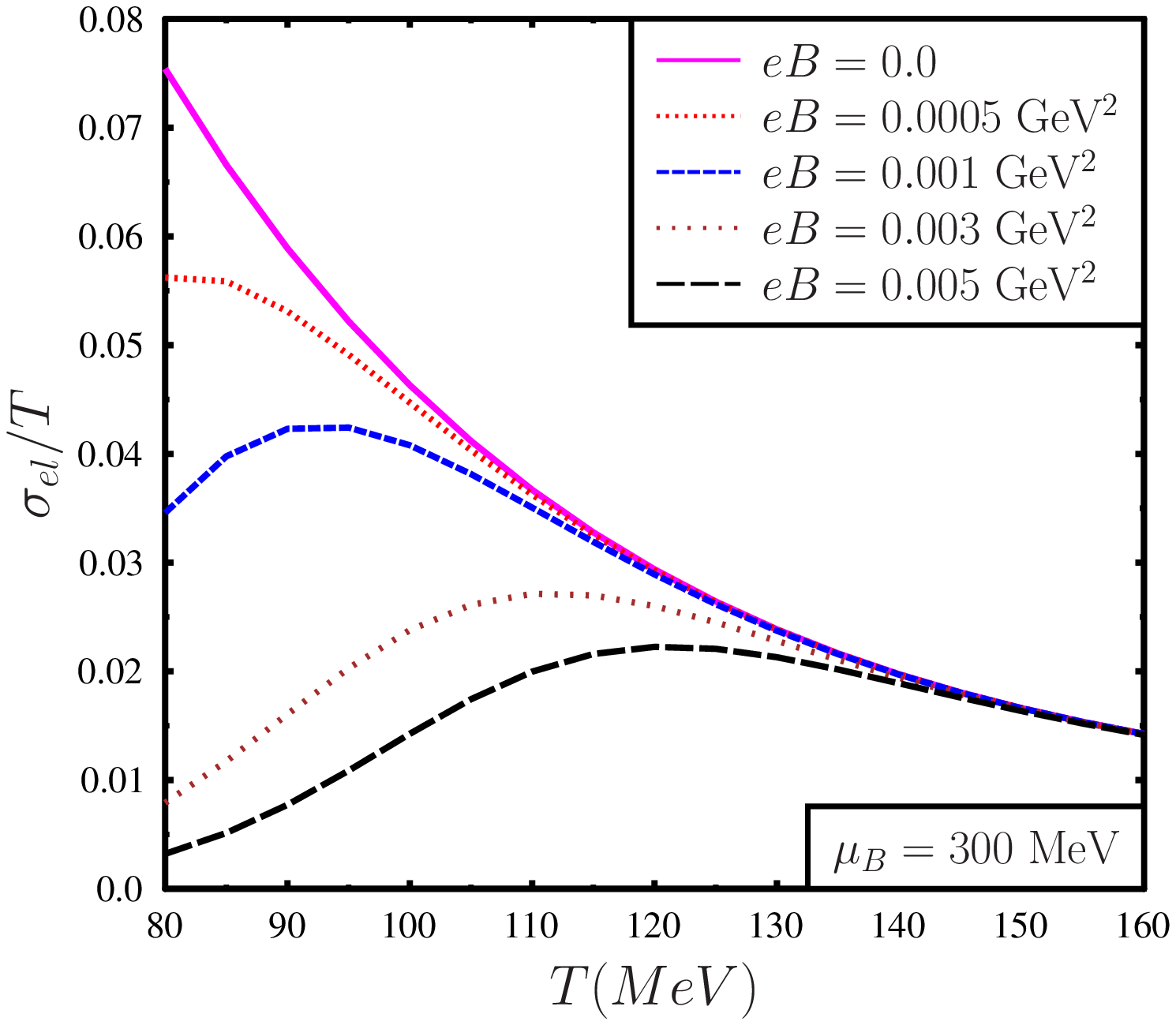}
    \end{minipage}~~
    \begin{minipage}{0.48\textwidth}
        \centering
        \includegraphics[width=1.2\linewidth]{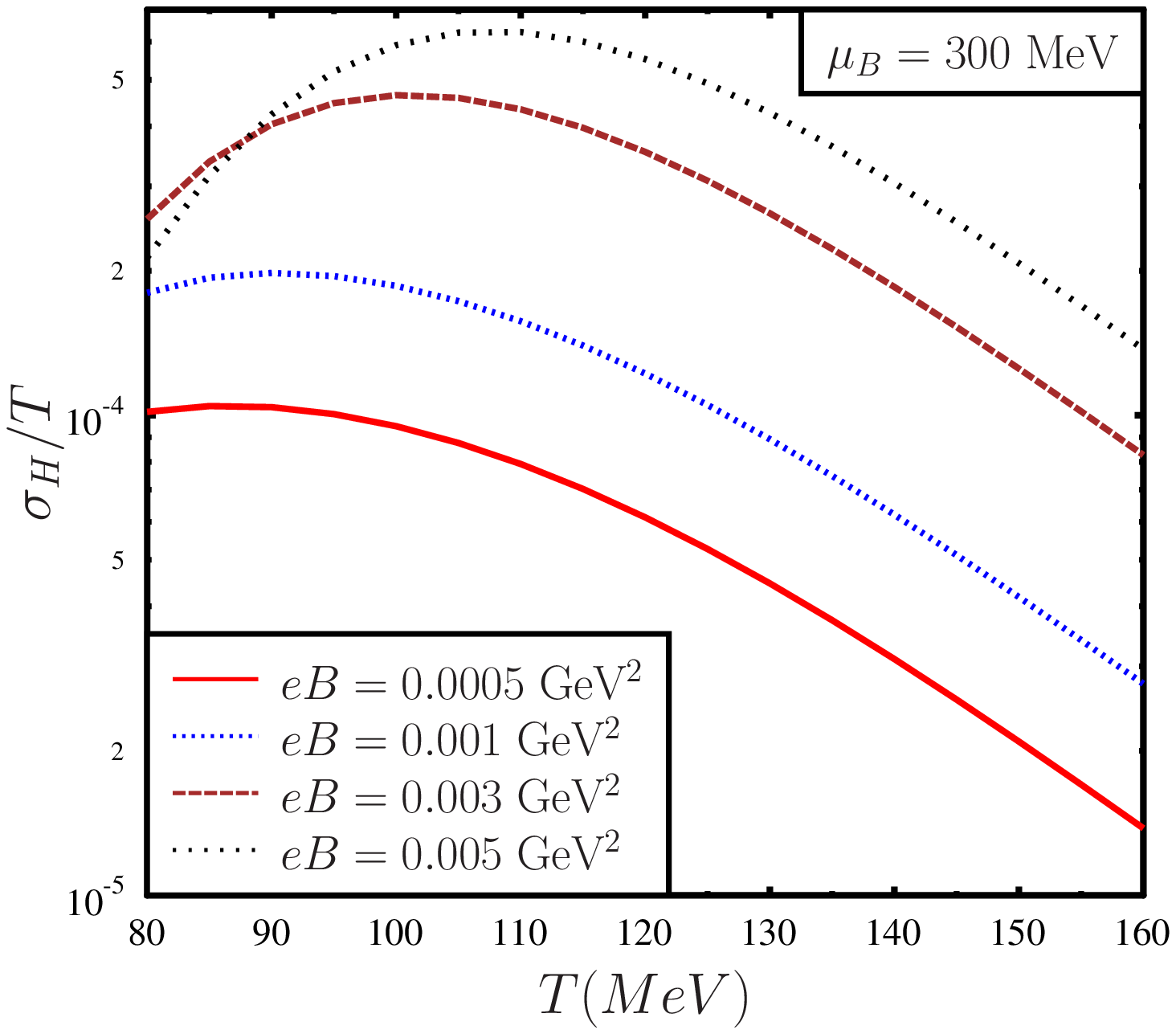}
    \end{minipage}
    \caption{Left plot: variation of normalized electrical conductivity, $\sigma_{el}/T\equiv \sum_a q_a^2 L_{1_a}/T$ with temperature for nonvanishing value of the magnetic field and baryon chemical potential. With magnetic field electrical conductivity generically decrease and with temperature $\sigma_{el}/T$ shows nonmonotonic behavior. In the high temperature range the effect of magnetic field is less significant on $\sigma_{el}/T$. Right plot:variation of normalized Hall conductivity $\sigma_{H}/T\equiv \sum_a q_a^2 L_{2_a}/T$ with temperature for nonvanishing magnetic field. With increasing magnetic field generically Hall conductivity increases for the range of magnetic field considered here.}
\label{fig6}
 \begin{minipage}{.48\textwidth}
        \centering
        \includegraphics[width=1.2\linewidth]{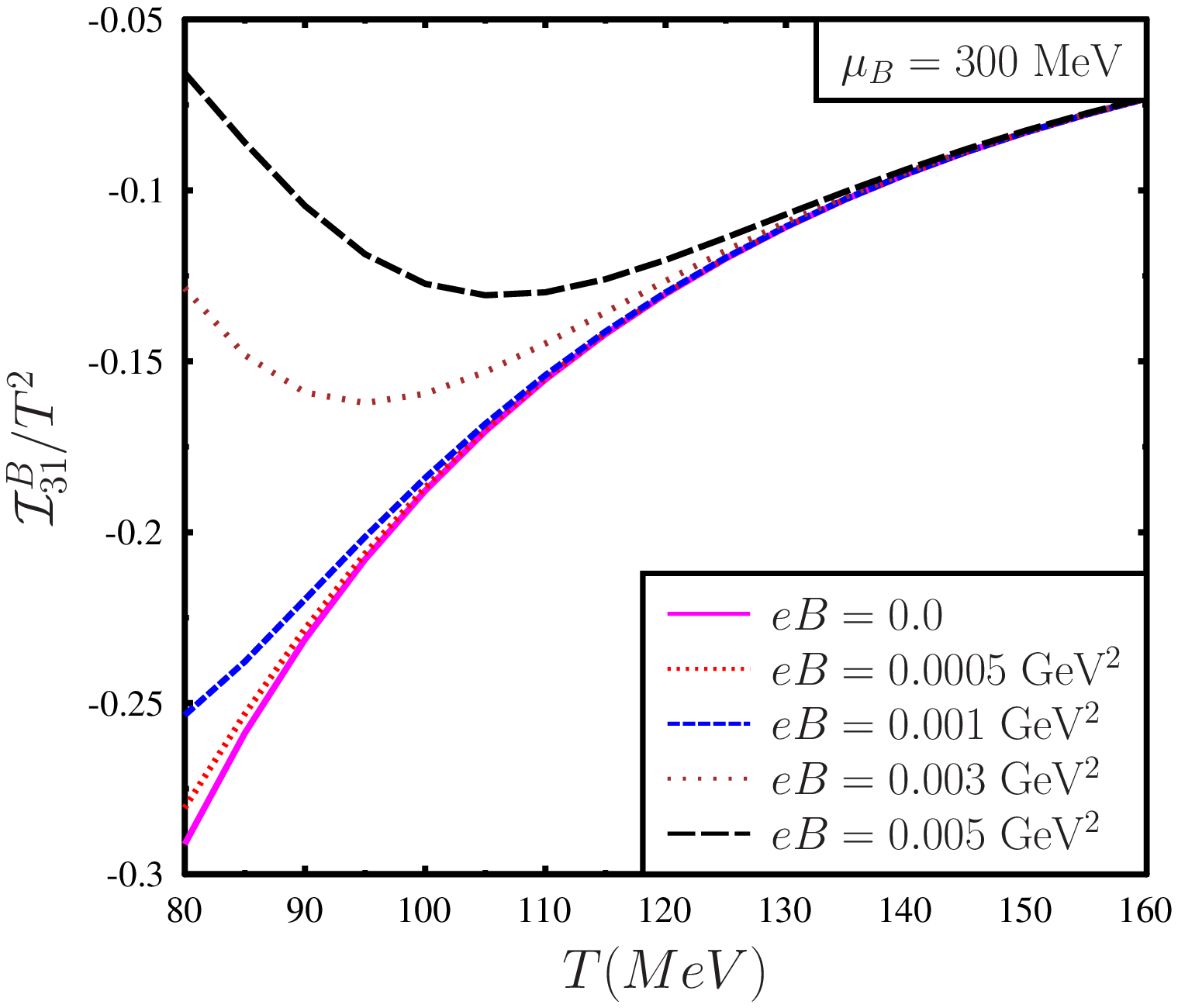}
    \end{minipage}~~
    \begin{minipage}{0.48\textwidth}
        \centering
        \includegraphics[width=1.2\linewidth]{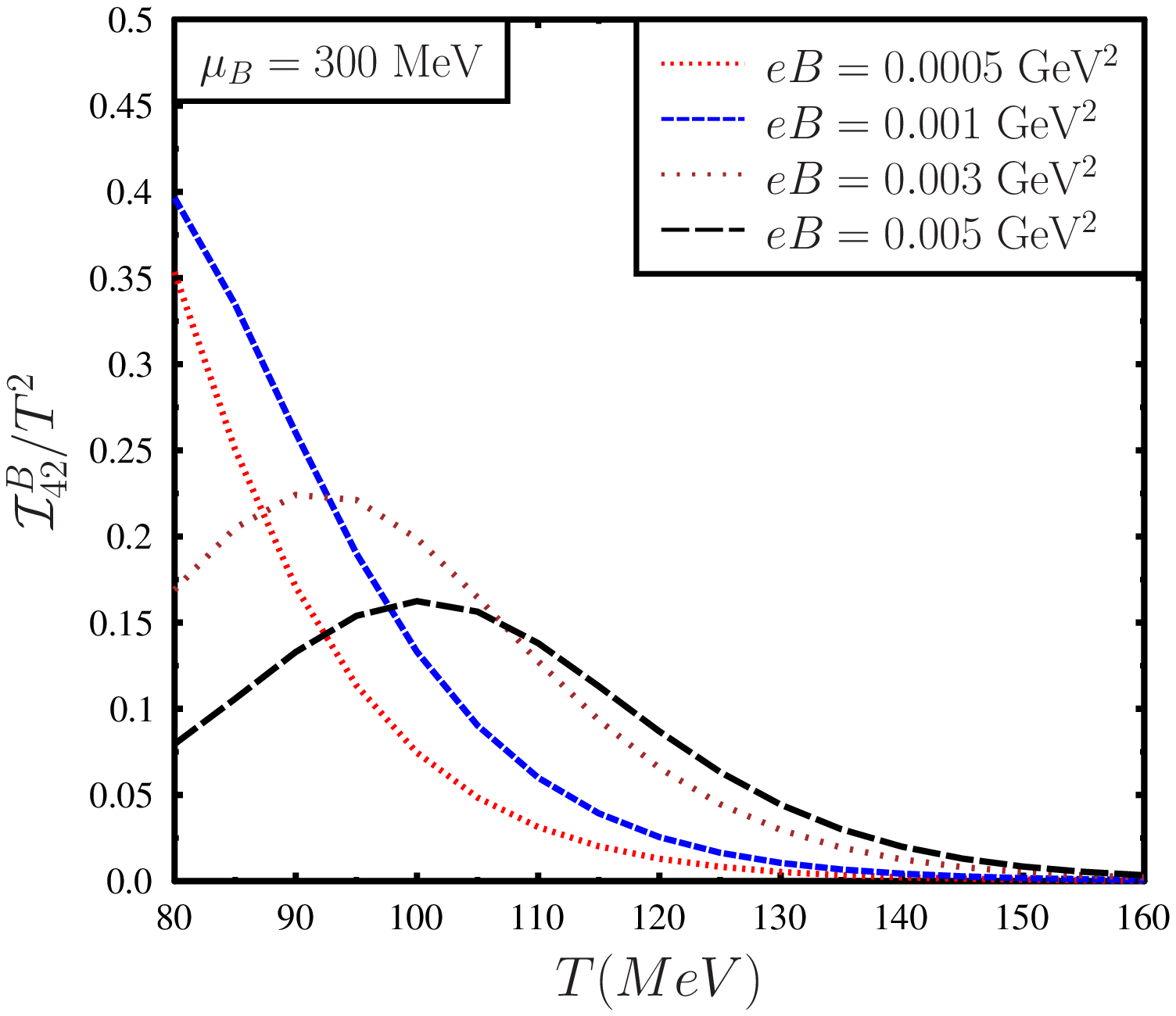}
    \end{minipage}
    \caption{Left plot: variation of $\mathcal{I}^B_{31}/T^2$ with temperature at finite baryon chemical potential for nonvanishing values of magnetic field. With increasing magnetic field
    $\mathcal{I}^B_{31}/T^2$ increase with respect to its value for vanishing magnetic field. Right plot: variation of $\mathcal{I}^B_{42}/T^2$ with temperature at finite baryon chemical potential for nonvanishing values of magnetic field. With magnetic field $\mathcal{I}^B_{42}/T^2$ first increases, then it decreases for higher values of magnetic field. Also for higher values of magnetic field it shows a nonmonotonic variation with temperature for the range of temperature considered here.}
\label{fig7}
\end{figure}

Next in Fig.\eqref{fig5} we show the variation of magneto-Seebeck coefficient ($S_B$) with temperature and baryon chemical potential for different values of the magnetic field. From this figure we can observe that with increasing magnetic field the magneto-Seebeck coefficient decreases in the low temperature range ($T\lesssim 120$ MeV). However in the high temperature range 
magnetic field does not affect significantly the Seebeck coefficient. Further with baryon chemical potential $S_B$ increases for the range of temperature and magnetic field considered in this investigation.
The variation of $S_B$ with magnetic field can be understood from Eq.\eqref{equnew49}. In this context we have shown the variation of $\sigma_{el}/T$ and $\sigma_{H}/T$ with magnetic field in Fig.\eqref{fig6}. From the left plot in Fig.\eqref{fig6} we can see that with increasing magnetic field, $\sigma_{el}/T$ decreases. However in the high temperature range the effect of magnetic field on $\sigma_{el}/T$ is less significant. The variation of $\sigma_{el}/T$ with magnetic field has been discussed in great length in Ref.\cite{ranjitaarpanhm1}. Basically in the low temperature range magnetic field affect the normalized electrical conductivity because in the low temperature range due to large value of the relaxation time $\omega_c\tau$ term in the expression of $\sigma_{el}$ becomes significant. With increasing magnetic field this factor of $\omega_c\tau$ is responsible for the decrease in $\sigma_{el}$. On the other hand in the high temperature range due to very small value of the relaxation time $\omega_c\tau$ term remains ineffective. Contrary to $\sigma_{el}/T$, the normalized Hall conductivity ($\sigma_{H}/T$) generically increases with magnetic field. For a detailed discussion on the behavior of $\sigma_{H}/T$ with magnetic field, see Ref.\cite{ranjitaarpanhm1}.  

Further in Fig.\eqref{fig7} we have shown the variation of $\mathcal{I}^B_{31}/T^2$ and $\mathcal{I}^B_{42}/T^2$ with magnetic field. With magnetic field $\mathcal{I}^B_{31}/T^2$ increases.
Also in the low temperature range the effect of magnetic field on $\mathcal{I}^B_{31}/T^2$ is significant and on the high temperature end due to small value of the relaxation time magnetic field does not affect $\mathcal{I}^B_{31}/T^2$ significantly. From the right plot in Fig.\eqref{fig7} we observe that with magnetic field 
first $\mathcal{I}^B_{42}/T^2$ increases and then with increasing magnetic field it decreases. 
From Fig.\eqref{fig6} it is clear that $\sigma_{el}/T$ is order of magnitude larger than $\sigma_{H}/T$. On the other hand the absolute values of $\mathcal{I}^B_{31}/T^2$ and $\mathcal{I}^B_{42}/T^2$ does not differ much, which can be observed in Fig.\eqref{fig7}. Thus for the situation when $\sigma_{el}/T$ is larger than $\sigma_{H}/T$
the variation of $S_B$ with temperature, chemical potential and magnetic field is predominantly determined by the factors, $\sigma_{el}/T$, $\mathcal{I}^B_{31}/T^2$ and under this condition $S_B\sim (\mathcal{I}^B_{31}/T^2)/(\sigma_{el}/T)$.
Hence the variation of $\mathcal{I}^B_{31}/T^2$ and $\sigma_{el}/T$ with magnetic field effectively give rise to the variation of $S_B$ with magnetic field as has been shown in the left plot in Fig.\eqref{fig5}.

\begin{figure}[!htb]
    \centering
    \begin{minipage}{.48\textwidth}
        \centering
        \includegraphics[width=1.2\linewidth]{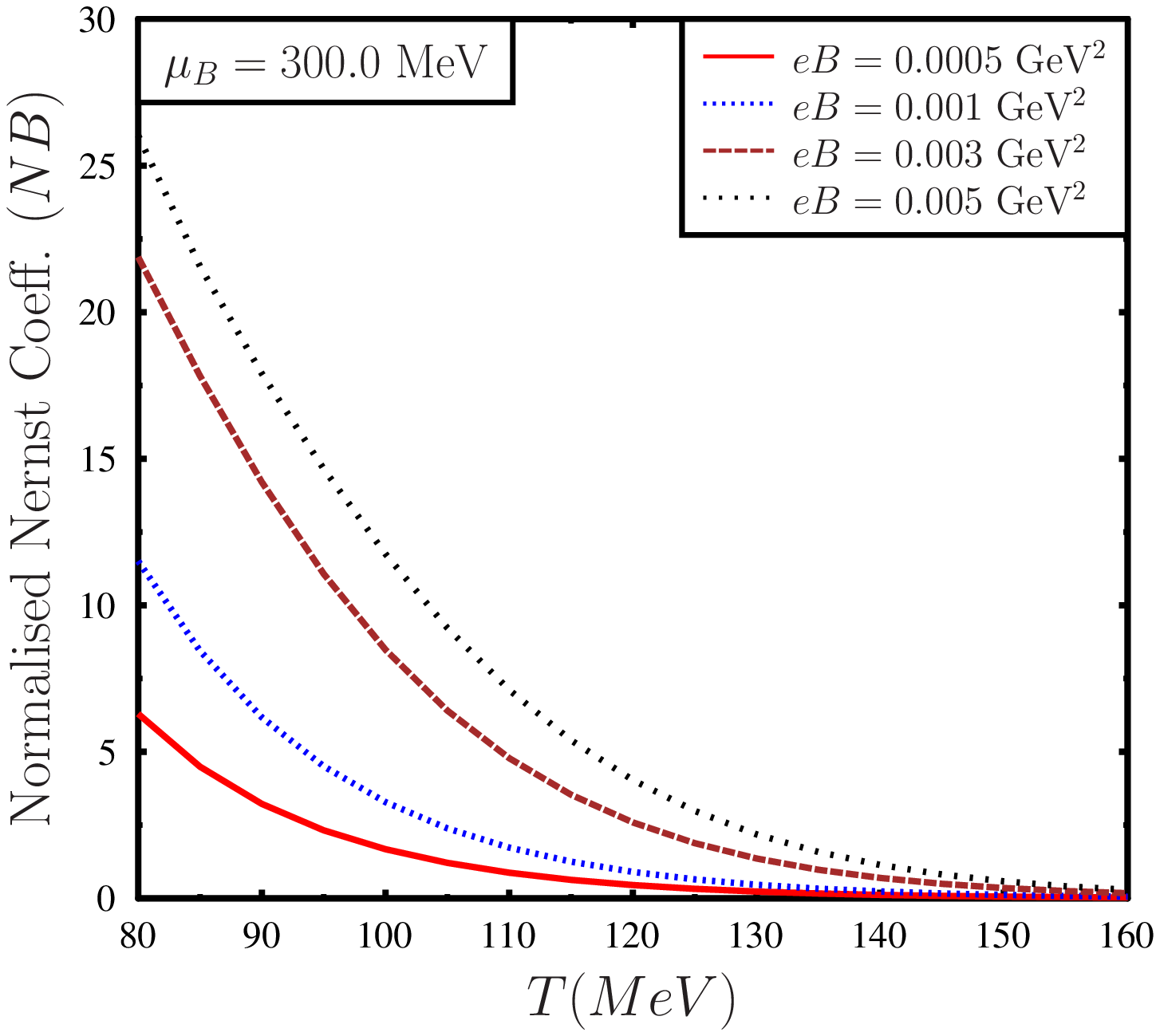}
    \end{minipage}
    \begin{minipage}{0.48\textwidth}
        \centering
        \includegraphics[width=1.2\linewidth]{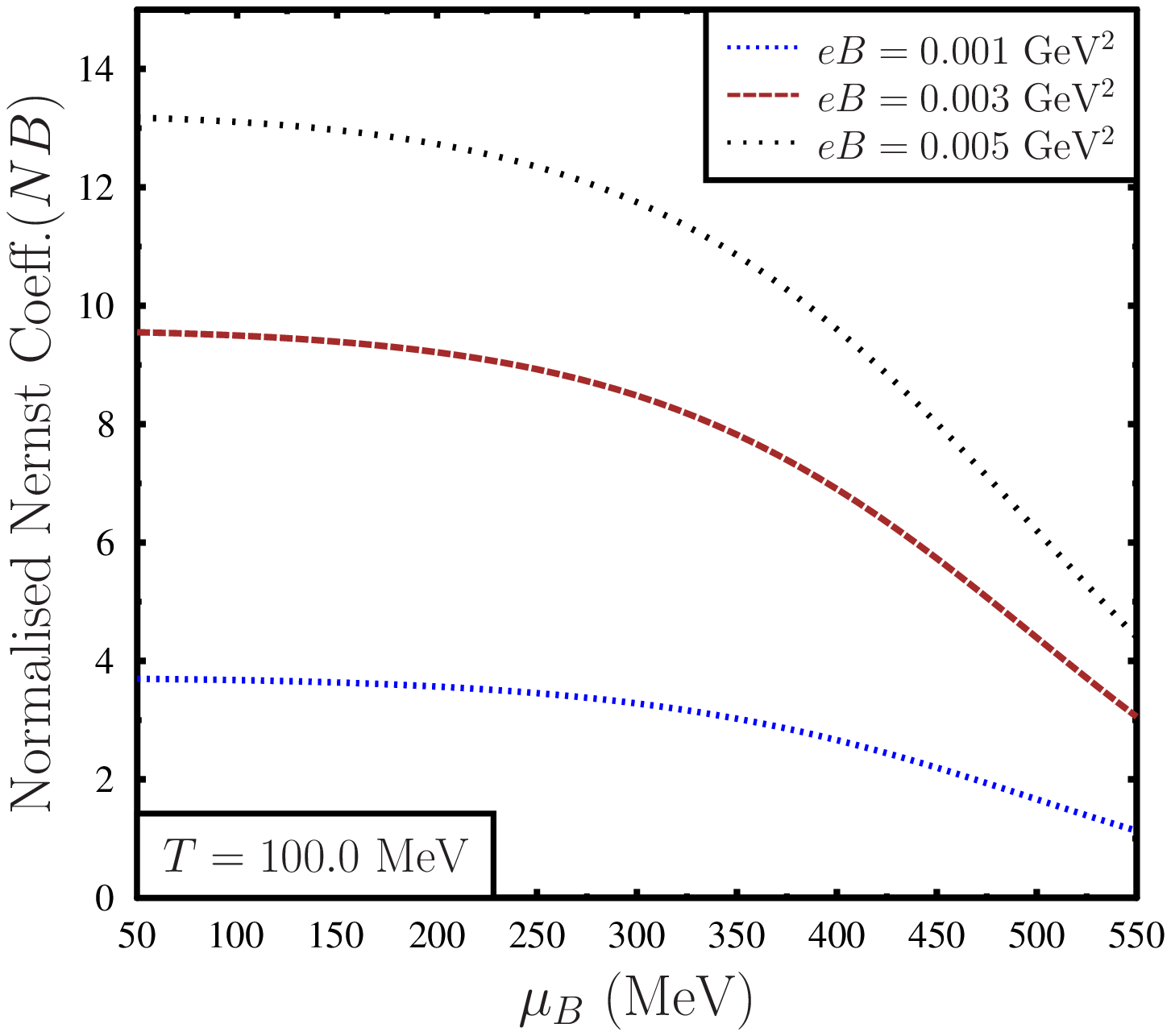}
    \end{minipage}
    \caption{Left plot: variation of normalised Nernst coefficient ($NB$) with temperature and magnetic field. With temperature $NB$ decrease and $NB$ increases with increasing magnetic field. However in the high temperature range the effect of magnetic field on $NB$ is less significant. Right plot: Variation of $NB$ with baryon chemical potential $\mu_B$ for nonvanishing magnetic field. With baryon chemical potential $NB$ decreases.}
\label{fig8} 
\end{figure}

Finally we discuss the variation of normalized Nernst coefficient $(NB)$ with temperature and baryon chemical potential for finite magnetic field. In Fig.\eqref{fig8} we show the variation of normalized Nernst coefficient $(NB)$ with temperature and baryon chemical. From Fig.\eqref{fig8} we can observe that with increasing temperature Nernst coefficient $(NB)$ decreases.
Due to the order of magnitude difference in the values of $\sigma_{el}/T$ and $\sigma_{H}/T$ 
the variation of normalized Nernst coefficient is
predominantly determined by $\sigma_{el}/T$ and
$\mathcal{I}^B_{42}/T^2$. Although $\mathcal{I}^B_{42}/T^2$ shows a nonmonotonic variation with magnetic field but the normalized Nernst coefficient does not show any nonmonotonic variation with magnetic field, probably because of the fact that with magnetic field $\sigma_{el}/T$ decrease. 
This decrease in $\sigma_{el}/T$ with magnetic field overwhelms the nonmonotonic variation of $\mathcal{I}^B_{42}/T^2$. Similarly with baryon chemical potential both $\mathcal{I}^B_{42}/T^2$ and $\sigma_{el}/T$ decreases as has been shown in Figs.\eqref{fig3} and \eqref{fig4}. But the rapid decrease of $\mathcal{I}^B_{42}/T^2$ relative to $\sigma_{el}/T$ possibly give rise to
the decreasing behavior of the normalized Nernst coefficient with baryon chemical potential.

\section{Conclusions}
In heavy-ion collision there is a gradient of temperature from central to peripheral region. This can to lead to a electric field and the relevant Seebeck coefficient as in the condensed matter systems. We had estimated this coefficient within HRG model in Ref.\cite{arpanhm}. In that investigation, however, the chemical potential was taken to be uniform. In the present work we have also included the gradient of the baryon chemical potential apart from a spatial gradient in temperature in the estimation of the thermoelectric coefficients. Conservation of momentum of the system alongwith the Gibbs-Duhem relation are used to relate to the gradient of baryon chemical potential to gradient in temperature. This leads to a modification of the expression for the Seebeck coefficient involving the enthalpy of the system compared to Ref.\cite{arpanhm}. Although because of the opposite charges of the mesons, mesonic Seebeck coefficient vanish, nonetheless mesons affect the Seebeck coefficient of the system significantly. This is because the mesonic contribution to the enthalpy of the system is dominant. This leads to a negative Seebeck coefficient of the hadronic system, unlike as in Ref.\cite{arpanhm} where the spatial gradient of the chemical potential was not taken into account leading to a positive Seebeck coefficient.

We have also considered the effects of magnetic field on the thermoelectric properties. Magnetic field leads to a Hall type current due to the Lorentz force. It is observed that the resulting magneto-Seebeck coefficient generically decrease with magnetic field for a given temperature and baryon chemical potential. However at higher temperature and/or chemical potential the magneto-Seebeck coefficient goes over to the Seebeck coefficient at vanishing magnetic field. 
The induced current perpendicular to the electric field is decided by the Nernst coefficient. It is also observed that the Nernst coefficient for the hadronic matter increases with magnetic field. However at high temperature or baryon chemical potential this approaches the vanishing value for the same at vanishing magnetic field. 

In the present investigation we have focused our attention for hadronic matter. It will definitely be interesting to investigate thermoelectric coefficients for the partonic matter as has been recently attempted in Ref.\cite{bkpatra}. In particular it will be interesting and important to investigate the effects of mean field and medium dependent masses on the thermoelectric properties of partonic matter.

\section*{Acknowledgements}
The idea of thermoelectric coefficient in the context of heavy-ion collision arose during a visit of one of the author’s (H.M.) to the research group of Prof. Ajit M. Srivastava at Institute of Physics Bhubaneswar.The authors would like to thank Ajit. M. Srivastava for originally suggesting the idea of 
thermoelectric coefficient in this context.
The authors would like to thank Prof. Jitesh R. Bhatt for useful discussions.
The authors would also like to thank Sabyasachi Ghosh, Abhishek Atreya, Aman Abhishek, Chowdhury Aminul Islam, Rajarshi Ray for many discussions on the topic of Seebeck coefficient during working group activities at WHEPP 2017, IISER Bhopal. The work of A.D. is supported by the Polish National Science Center Grants No. 2018/30/E/ST2/00432.

%\newpage


\begin{thebibliography}{99}
\bibitem{HeinzSnellings2013}
U. W. Heinz and R. Snellings, Annu. Rev. Nucl. Part. Sci. {\bf 63},
123-151 (2013).
\bibitem{RomatschkeRomatschke}
P. Romatschke and U. Romatschke, Phys.Rev.Lett. {\bf 99},172301 (2007).
\bibitem{KSS}
P.K. Kovtun, D.T.Son and A.O.Starinets, Phys.Rev.Lett. {\bf 94},111601 (2005).
\bibitem{gavin1985}
S. Gavin, Nucl. Phys. {\bf A 435}, 826 (1985).
\bibitem{kajantie1985}
A. Hosoya, K. Kajantie, Nucl. Phys. {\bf B250}, 666 (1985).
\bibitem{DobadoTorres2012}
A.Dobado and J. M. Torres-Rincon, Phys. Rev.{\bf D86}, 074021 (2012).
\bibitem{sasakiRedlich2009}
C. Sasaki and K.Redlich, Phys. Rev.{\bf C79}, 055207 (2009).
\bibitem{sasakiRedlich2010}
C. Sasaki and K.Redlich, Nucl. Phys.{\bf A832}, 62 (2010).
\bibitem{KarschKharzeevTuchin2008}
F. Karsch, D. Kharzeev, and K. Tuchin, Phys. Lett. {\bf B663}, 217 (2008).
\bibitem{FinazzoRougemont2015}
S. I. Finazzo, R. Rougemont, H. Marrochio, J. Noronha, JHEP 1502, 051 (2015).
\bibitem{WiranataPrakash2009}
A. Wiranata and M. Prakash, Nucl. Phys.{\bf A830}, 219C-222C (2009) 
\bibitem{JeonYaffe1996}
S. Jeon and L. Yaffe, Phys.Rev.{\bf D53}, 5799-5809 (1996).
\bibitem{mclerran2008}
D. E. Kharzeev, L. D. McLerran, H. J. Warringa, Nucl. Phys. {\bf A 803}, 227 (2008).
\bibitem{skokov}
V. Skokov, A. Yu. Illarionov, V. Toneev, Int. J. Mod. Phys. {\bf A 24}, 5925 (2009).

\bibitem{MHD1}
G. Inghirami et al., Eur. Phys. J. C 76,659 (2016).
\bibitem{MHDajit}
A. Das, S.S. Dave, P.S. Saumia, A.M. Srivastava, Phys.Rev.{\bf C96}, 034902 (2017).
\bibitem{TuchinMHD}
K.Tuchin, Phys.Rev.{\bf C83}, 017901 (2011); Phys. Rev.{\bf C82}, 034904 (2010).
\bibitem{MoritzGreif} 
M. Greif, C. Greiner, and G. S. Denicol, Phys.Rev. {\bf D93} (2016) no.9, 096012.
\bibitem{electricalcond1}
 M. Greif, I. Bouras, C. Greiner, and Z. Xu, Phys. Rev. {\bf D90}, 094014 (2014).
\bibitem{electricalcond2} 
 A. Puglisi, S. Plumari, and V. Greco, arXiv:1407.2559.
 \bibitem{electricalcond3}
 A. Puglisi, S. Plumari, and V. Greco, Phys. Rev. {\bf D90}, 114009 (2014).
 \bibitem{electricalcond4}
  W. Cassing, O. Linnyk, T. Steinert, and V. Ozvenchuk, Physical Review Letters 110, 182301 (2013).
  \bibitem{electricalcond5}
  T. Steinert and W. Cassing, Physical Review C 89, 035203 (2014).
  \bibitem{electricalcond6}
   G. Aarts, C. Allton, A. Amato, P. Giudice, S. Hands, and J.-I. Skullerud, JHEP 02, 186 (2015).
  \bibitem{electricalcond7}
   G. Aarts, C. Allton, J. Foley, S. Hands, and S. Kim, Physical Review Letters 99, 022002 (2007).
   \bibitem{electricalcond8}
    A. Amato, G. Aarts, C. Allton, P. Giudice, S. Hands,  and J.-I. Skullerud, Physical Review Letters 111, 172001 (2013).
 \bibitem{electricalcond9}
  S. Gupta, Physics Letters B 597, 57 (2004).
  \bibitem{electricalcond10}
   Y. Burnier and M. Laine, The European Physical Journal C 72, 1902 (2012).
   \bibitem{electricalcond11}
    H.-T. Ding, A. Francis, O. Kaczmarek, F. Karsch, E. Laermann, and W. Soeldner, Physical Review D 83, 034504 (2011).
\bibitem{electricalcond12}
 O. Kaczmarek and M. M̈uller, PoS LATTICE2013 , 175 (2014)
\bibitem{electricalcond13}
 S.-X. Qin, Phys. Lett. B742, 358 (2015).
\bibitem{electricalcond14}
 R. Marty, E. Bratkovskaya, W. Cassing, J. Aichelin, and H. Berrehrah, Phys. Rev. C88, 045204 (2013)
 \bibitem{electricalcond15}
  D. Fern ́andez-Fraile and A. Gomez Nicola, Physical Review D 73, 045025 (2006). 
\bibitem{kharzeevbook}
``Strongly interacting Matter in Magnetic field'', edited by D. Kharzeev, K. Landsteiner, A. Schmitt 
and H. Yee, Lecture Notes in Physics vol 871, Springer-Verlag Berlin Heidelberg 2013. 

  
  
  \bibitem{danicol2018}
 M. Greif, J. A. Fotakis, G. S. Denicol, C. Greiner, Phys. Rev. Lett. {\bf 120}, 242301 (2018). 
 \bibitem{PrakashVenu}
 M. Prakash, M. Prakash, R. Venugopalan and G. Welke, Phys.Rept.{\bf 227}, 321-366 (1993).
\bibitem{WiranataPrakash2012}
A. Wiranata and Madappa Prakash, Phys. Rev.{\bf C85}, 054908 (2012).
\bibitem{KapustaChakraborty2011}
 P. Chakraborty and J.I. Kapusta Phys. Rev.{\bf C83}, 014906 (2011).
\bibitem{Toneev2010}
 A.S. Khvorostukhin, V.D. Toneev, D.N. Voskresensky, Nucl. Phys. {\bf A845}, 106 (2010).
\bibitem{Plumari2012}
S.Plumari, A. Paglisi, F. Scardina and V. Greco,Phys. Rev.{\bf C86}, 054902 (2012).
 \bibitem{Gorenstein2008}
 M. I. Gorenstein, M. Hauer, O. N. Moroz, Phys.Rev.{\bf C77}, 024911 (2008).
\bibitem{Greiner2012}
 J. Noronha-Hostler, J. Noronha and C. Greiner , Phys. Rev. {\bf C86}, 024913 (2012).
\bibitem{TiwariSrivastava2012}
 S.K. Tiwari, P.K. Srivastava, C.P. Singh, Phys.Rev. {\bf C85}, 014908 (2012).
\bibitem{GhoshMajumder2013}
 S. Ghosh, A. Lahiri, S. Majumder, R. Ray, S. K. Ghosh, Phys. Rev. {\bf C88}, 068201 (2013).
\bibitem{Weise2015}
R. Lang, N. Kaiser, and W. Weise, Eur. Phys. J. {\bf A51}, 127 (2015).
\bibitem{GhoshSarkar2014}
 S. Ghosh, G. Krein, S. Sarkar, Phys.Rev. {\bf C89}, 045201 (2014).
\bibitem{WiranataKoch} 
A. Wiranata, V. Koch and M. Prakash, X.N. Wang, J.Phys.Conf.Ser.{\bf 509}, 012049 (2014). 
\bibitem{WiranataPrakashChakrabarty2012}
 A. Wiranata, M. Prakash and P. Chakraborty, Central Eur.J.Phys. {\bf 10}, 1349-1351 (2012).
\bibitem{Wahba2010}
 A. Tawfik and M. Wahba, Ann. Phys. {\bf 522}, 849-856 (2010).
 
 
\bibitem{Greiner2009}
J. Noronha-Hostler,J. Noronha and C. Greiner, Phys. Rev. Lett.{\bf103}, 172302 (2009).
\bibitem{KadamHM2015}
G. Kadam, H. Mishra, Nuclear Physics {\bf A934}, 133147  (2015).
\bibitem{Kadam2015}
G. Kadam, Mod.Phys.Lett. {\bf A30}, no.10, 1550031  (2015).
\bibitem{Ghoshijmp2014}
S. Ghosh, Int. J. Mod. Phys. {\bf A29}, 1450054 (2014).
\bibitem{Demir2014}
N. Demir and A. Wiranata, J.Phys.Conf.Ser.{\bf 535}, 012018 (2014). 
\bibitem{Ghosh2014}
S. Ghosh, Phys. Rev. {\bf C90}, 025202 (2014).
\bibitem{smash}
J.-B. Rose, J. M. Torres-Rincon, A. Schäfer, D. R. Oliinychenko, and H. Petersen, Phys. Rev. C 97, 055204 (2018).
\bibitem{bamps}
C. Wesp, A. El, F. Reining, Z. Xu, I. Bouras, and C. Greiner, Phys. Rev. C 84, 054911 (2011).
\bibitem{bamps2}
Moritz Greif, Ioannis Bouras, Carsten Greiner, and Zhe Xu, Phys. Rev. D 90, 094014 (2014).
\bibitem{urqmd1}
S. A. Bass et. al, Prog. Part. Nucl. Phys. {\bf 41}, 225 (1998).
\bibitem{GURUHM2015}
G. Kadam, H. Mishra, Phys. Rev. {\bf C92}, 035203 (2015).
\bibitem{ranjitahm}
R. K. Mohapatra, H. Mishra, S. Dash, B. K. Nandi, arXiv:1901.07238.
\bibitem{amanhm1}
P. Singha, A. Abhishek, G. Kadam, S. Ghosh, H. Mishra, J. Phys. {\bf G46}, 015201.
\bibitem{amanhm2}
A. Abhishek, H. Mishra, S. Ghosh, Phys. Rev. {\bf D97}, 014005 (2018).
\bibitem{arpanhm}
J. R. Bhatt, A. Das, H. Mishra, Phys. Rev. {\bf D99}, 014015 (2019).
\bibitem{ranjitaarpanhm1}
A. Das, H. Mishra, R. K. Mohapatra,
Phys.Rev. D99, 094031 (2019).
\bibitem{ranjitaarpanhm2}
A. Das, H. Mishra, R. K. Mohapatra,
Phys.Rev. D100, 114004 (2019).
\bibitem{ranjitaarpanhm3}
A. Das, H. Mishra, R. K. Mohapatra, 
arXiv:1907.05298, Phys.Rev. D101, 034027 (2020).
\bibitem{jayantadey1}
J. Dey, S. Satapathy, P. Murmu, S. Ghosh,
arXiv:1907.11164.
\bibitem{feng2017}
B. Feng, Phys. Rev. {\bf D96}, 036009 (2017).


\bibitem{danicol2014}
G.S. Denicol, H. Niemi, I. Bouras E. Molnar , Z. Xu , D.H. Rischke, C. Greiner ,Phys. Rev. {\bf D 89}, 074005 (2014).
\bibitem{Kapusta2012}
J.I. Kapusta and J.M. Torres-Rincon,Phys. Rev. {\bf C86}, 054911 (2012).
\bibitem{callen}
H.  B.  Callen, Thermodynamics (Wiley,  New York,  1960).
\bibitem{sofo}
T.J.Scheidemantel, C. Ambrosch-Draxi, T.Thonhauser, J.V.Badding and J.O.Sofo, Phys.Rev.{\bf B68},125210 (2003).
\bibitem{conseeb1}
P. Ao, arXiv:cond-mat/9505002;  M. Matusiak, K. Rogacki, T. Wolf,  Phys. Rev. B 97, 220501(R) (2018); 
M. K. Hooda, C. S. Yadav, arXiv:1704.07194;  O. Cyr-Choiniere et. al.,  Phys. Rev. X 7, 031042 (2017);
 L. P. Gaudart, D. Berardan, J. Bobroff, N. Dragoe,  Phys. Stat. Sol. (RRL) 2, No. 4, 185-187 (2008).
\bibitem{conseeb2}
 S. Sergeenkov, JETP Letters 67 (1998) 650-655. 
\bibitem{conseeb3}
 M. Wysokinski, J. Spalek, J. Appl. Phys. 113, 163905 (2013).
\bibitem{conseeb4}
 K. P. Wojcik, I. Weymann, Phys. Rev. B 89, 165303 (2014). 
\bibitem{conseeb5}
 Kangjun Seo, Sumanta Tewari, Physical Review B 90, 174503 (2014).
\bibitem{conseeb6}
 P. Dutta, A. Saha, A. M. Jayannavar,  Phys. Rev. B 96, 115404 (2017);  S. Kolenda, M. J. Wolf, D. Beckmann
  Phys. Rev. Lett. 116, 097001 (2016).
\bibitem{conseeb7}
M. Shahbazi, C. Bourbonnais, Phys. Rev. B 94, 195153 (2016).
\bibitem{linearizedBoltzmann}
A. Cantarero and F. X. Alvarez, ``Thermoelectric Effects: Semiclassical and Quantum Approaches
from the Boltzmann Transport Equation'', published in Lect. Notes in Nanoscale Science and 
Technology, vol 16 ``Nanoscale Thermoelectrics'', edited by X. Wang and Z. M. Wang. 
\bibitem{thermoelectrics}
G.S.Nolas, J. Sharp and H. J. Goldsmid, ``Thermoelectrics: Basic Principles
and New Materials Developments'', Springer series in Materials Science, vol 45.
\bibitem{onsager1}
R. Wolfe, G.E. Smith, Phys. Rev. 129, 1086 (1963).

\bibitem{fairref}
See,``https://www.gsi.de/en/researchaccelerators/fair.htm'.
\bibitem{nica}
See,``http://nica.jinr.ru''.


%\bibitem{tuchin1}
%K. Tuchin, Phys. Rev. {\bf C88}, 024911 (2013).
%\bibitem{tuchin2}
%K. Tuchin, Adv. High Energy Phys. 2013, 490495 (2013). 
%\bibitem{semiconductor}
%``Basic Semiconductor Physics'', C. Hamaguchi, Spinger, Spinger-Verlag Berlin Heidelberg 2001, 2010, DOI: 10.1007/978-3-642-03303-2
%\bibitem{pairplasma1}
%A.Kandus, C. G. Tsagas, Mon. Not. R. Astron. Soc. {\bf 385}, 883-892 (2008).
%\bibitem{pairplasma2}
%E. G. Blackman, G. B. Field, Phys. Rev. Lett. {\bf 71}, 3481 (1993).
%\bibitem{pairplasma3}
%N. Bessho, A. Bhattacharjee, Physics of Plasmas {\bf 14}, 056503 (2007).


\bibitem{HRG1}
P. Braun-Munzinger, K. Redlich, J. Stachel, nucl-th/0304013.
\bibitem{HRG2}
 A. Andronic, P. Braun-Munzinger, J. Stachel, Nucl. Phys. {\bf A772}, 167 (2006).
 \bibitem{HRG3}
   P. Braun-Munzinger, D. Magestro, K. Redlich, J. Stachel, Phys. Lett. {\bf B518}, 41 (2001);
  Cleymans, K. Redlich, Phys. Rev.{\bf C60}, 054908 (1999);
   F. Becattini, et al., Phys. Rev.{\bf C64}, 024901 (2001);
 Cleymans, B. Kampfer, M. Kaneta, S. Wheaton, N. Xu, Phys. Rev.{\bf C71}, 054901 (2005);
 A. Andronic, P. Braun-Munzinger, J. Stachel, Phys. Lett. {\bf B673}, 14 (2009).
 
 \bibitem{HRG4}
 R. Dashen, S. Ma, and H. J. Bernstein, Phys. Rev. {\bf187}, 345 (1969).

 \bibitem{HRG5}
R. Dashen and R. Rajaraman, Phys.Rev. {\bf D10}, 694 (1974).

\bibitem{thermodynamicsHRG1}
F. Karsch, K. Redlich, A. Tawfik, Phys.Lett. B571, 67-74 (2003).
\bibitem{thermodynamicsHRG2}
P. Braun-Munzinger, V. Koch, T. Schafer, J. Stachel,  Phys.Rept. 621, 76 (2016).




\bibitem{hrgfluc3}
M. Nahrgang, M. Bluhm, P. Alba, R. Bellwied, C. Ratti, Eur.Phys.J. C75, no.12, 573 (2015).
\bibitem{hrgfluc4}
A. Bhattacharyya, S. Das, S. K. Ghosh, R. Ray, S. Samanta,
Phys.Rev. {\bf C}90, no.3, 034909 (2014).
\bibitem{hrgfluc5}
P. Garg, D.K. Mishra, P.K. Netrakanti, B. Mohanty, A.K. Mohanty,
B.K. Singh, N. Xu,  Phys.Lett. B726, 691-696 (2013).
\bibitem{hrgfluc6}
A. Bazavov et al. Phys.Rev. {\bf D} 86, 034509 (2012).
\bibitem{hrgfluc7}
V.V. Begun, M. I. Gorenstein, M. Hauer, V.P. Konchakovski, O.S. Zozulya,
Phys.Rev. {\bf C} 74, 044903 (2006).
\bibitem{ranjita2019}
R. K. Mohapatra, Phys. Rev. C99, 024902 (2019).
\bibitem{stockerRischke}
D.H.Rischke, M.I.Gorenstein, H.Stocker, W.Greiner; Z. Phys. C 51,485-489 (1991)
\bibitem{magSeeb1}
 Y. Hasegawa, T. Komine, Y. Ishikawa, A. Suzuki and H. Shirai, Jpn. J. Appl. Phys. 43, 35 (2004).
\bibitem{HRGMuller}
A. Majumder and B. Muller, Phys. Rev. Lett {\bf 105},252002 (2010).


\bibitem{lataguruhm}
G. Kadam, H. Mishra, L. Thakur, Phys. Rev. {\bf D 98}, 114001 (2018). 

 
\bibitem{paramitahm2016}
P. Deb, G. Kadam, H. Mishra, Phys. Rev. {\bf D 94},094002 (2016).
\bibitem{gondologelmini}
P. Gondolo and G. Gelmini, Nucl. Phys. {\bf B360}, 145 (1991).
\bibitem{pdg}
C. Amsler et al.[Particle Data Group], Phys. Lett. {\bf B 667}, 1 (2008).
\bibitem{kapustaAlbr}
M. Albright, J. Kapusta, C. Young, Phys.Rev. C {\bf 90} no.2, 024915 (2014). 
\bibitem{hmgururadius1}
P. Braun-Munzinger, I. Heppe, J. Stachel, Phys. Lett. {\bf B465}, 15 (1999).
\bibitem{bkpatra}
D. Dey, B. K. Patra,  arXiv: 2004.03149. 
\end{thebibliography}
\end{document}